\documentclass{aastex63}

\usepackage{booktabs}
\usepackage{chemfig}
\usepackage{mathtools}
\usepackage{amsmath}
\usepackage{multirow}

\shorttitle{The interstellar heritage of comets}
\shortauthors{Willacy et al.}

\begin{document}
\title{The interstellar heritage of comets}

\correspondingauthor{Karen Willacy}
\author[0000-0001-6124-5974]{Karen Willacy}
\email{karen.willacy@jpl.nasa.gov}
\affil{Jet Propulsion Laboratory, California Institute of Technology, MS 169-506, 
4800 Oak Grove Drive,
Pasadena, CA 91109, USA}

\author[0000-0001-7031-8039]{Liton Majumdar}
\affil{Exoplanets and Planetary Formation Group, School of Earth and Planetary Sciences, National Institute of Science Education and Research, Jatni 752050, Odisha, India}
\affil{Homi Bhabha National Institute, Training School Complex, Anushaktinagar, Mumbai 400094, India}

\author[0000-0002-6391-4817]{Boncho Bonev}
%\email{bonev@american.edu}
\affil{Department of Physics, American University, Washington D.C., USA}

\author[0000-0003-0142-5265]{Erika Gibb}
%\email{gibbe@umsl.edu}
\affil{Department of Mathematics, Physics, Astronomy and Statistics, University of Missouri-St Louis, St Louis, MO, USA}

\author[0000-0002-8379-7304]{Neil Dello Russo}
%\email{neil.dello.russo@jhuapl.edu}
\affil{Johns Hopkins Applied 
Physics Laboratory, 11100 Johns Hopkins Road, Laurel, MD 20723, USA}

\author[0000-0001-8843-7511]{Michael DiSanti}
%\email{michael.a.disanti@nasa.gov}
\affil{Solar System Exploration Division, Planetary Science Laboratory Code 693, NASA/Goddard Space Flight Center, Greenbelt, MD, USA}
\affil{Goddard Center for Astrobiology, NASA/Goddard Space Flight Center, Greenbelt, MD, USA}

\author[0000-0002-8227-9564]{Ronald J. Vervack, Jr.}
%\email{ron.vervack@jhuapl.edu}
 \affil{Johns Hopkins Applied Physics Laboratory, 11100 Johns Hopkins Road, Laurel, MD 20723, USA}

\author[0000-0002-6006-9574]{Nathan X. Roth}
%\email{nathan.x.roth@nasa.gov}
\affil{Department of Physics, American University, Washington, DC 20016, USA} \affil{Solar System Exploration Division, Astrochemistry Laboratory Code 691, NASA Goddard Space Flight Center, 8800 Greenbelt Road, Greenbelt, MD 20771, USA}

\begin{abstract}

Comets have similar compositions to interstellar medium ices, suggesting at least some of their molecules maybe inherited from an earlier stage of evolution. To investigate the degree to which this might have occurred we compare the composition of individual comets to that of the well-studied protostellar region IRAS 16293-2422B. We show that the observed molecular abundance ratios in several comets correlate well with those observed in the protostellar source. However, this does not necessarily mean that the cometary abundances are identical to protostellar. We find the abundance ratios of many molecules present in comets are enhanced compared to their protostellar counterparts. For COH-molecules, the data suggest higher abundances relative to methanol of more complex species, e.g. HCOOH, CH$_3$CHO, and HCOOCH$_3$, are found in comets. For N-bearing molecules, the ratio of nitriles relative to CH$_3$CN -- HC$_3$N/CH$_3$CN and HCN/CH$_3$CN -- tend to be enhanced. The abundances of cometary SO and SO$_2$ relative to H$_2$S are enhanced, whereas OCS/H$_2$S is reduced. Using a subset of comets with a common set of observed molecules we suggest a possible means of determining the relative degree to which they retain interstellar ices. This analysis suggests that over 84\% of COH-bearing molecules can be explained by the protostellar composition. The possible fraction inherited from the protostellar region is lower for N-molecules at only 26--74\%. While this is still speculative, especially since few comets have large numbers of observed molecules, it provides a possible route for determining the relative degree to which comets contain disk-processed material.
\end{abstract}


\section{Introduction}

Comets are among the least processed objects in the solar system, preserving volatiles that  provide valuable insights into the physical and chemical conditions at the time and location of their formation \citep{bergin23}. Once formed comets were scattered into  two main reservoirs -- the Oort cloud and the Kuiper belt -- where they remained at very cold temperatures until being scattered into the inner solar system.  Despite the potential for an unknown degree of post-formation processing, comets provide a unique opportunity to investigate volatile chemistry during the early development of the solar system. Indeed it is possible that their composition could reflect an even earlier stage of evolution, originating at least in part in the parent molecular cloud, where similar molecules are observed as ices. 

Several previous authors have explored links between cometary and interstellar molecular abundance ratios.  \cite{bm00} compared the volatiles observed in comet Hale-Bopp with gas phase molecular abundances observed in hot cores where the temperatures are high enough for ices to desorb, so the gas-phase abundances reflect those of ices accumulated in an earlier cold, dark core phase. They found a strong correlation between the abundances of COH- and N-bearing molecules relative to CH$_3$OH in the comet and hot cores, suggesting a link between the two, with little processing required in the disk to explain the cometary abundances. More recently, \cite{droz19} (hereafter D19) compared the relative abundances of molecules observed in IRAS 16293-2422B with those detected in comet 67P/Churyumov-Gerasimenko and found a strong correlation between the abundance ratios in the two sources, further supporting the connection between interstellar and cometary ices. 

A larger number of comets was considered by \cite{bianchi19}, who compared the abundances of complex organic molecules (COMs) in the hot corino of SVS13-A with those detected in the comets Hale-Bopp, Lemmon, Lovejoy and 67P. The abundance ratios relative to CH$_3$OH of methyl formate (HCOOCH$_3$) and ethanol (C$_2$H$_5$OH) in these comets are within a factor of 10 of the detections in SVS13-A, whereas formamide (NH$_2$CHO) and acetaldehyde (CH$_3$CHO) are enhanced in the comets, suggesting some inheritance of COMs from an earlier stage of evolution, but with additional processing required to account for the other COMs.

A different approach was taken by \cite{lippi24} who carried out a statistical comparison between the abundances of NH$_3$, CH$_3$OH and H$_2$CO in comets and those in protostellar disks. They found similar abundance ratios for CH$_3$OH/H$_2$CO and NH$_3$/CH$_3$OH suggesting that cometary compositions could have been inherited from the early stages of the formation of the solar system.

These papers suggest that cometary ices are inherited, at least in part, from the interstellar medium. However, comets vary greatly in composition. Some show enhanced organic abundances (e.g., \cite{rubin19b} and references therein), whereas others are depleted in organics, e.g. C/1999 S4 (LiNEAR) \citep{mumma01} and 73P/SW3 \citep{villanueva06, dr07}. \cite{dr16} demonstrate that there is at least an order of magnitude variation in volatile abundance ratios relative to water, and that Jupiter family comets tend to be more depleted in volatiles than Oort cloud comets. While some ices may be inherited from the interstellar medium, some degree of processing in the protosolar disk would seem to be required, with the amount of processing varying among comets.

Here we build on the previous work by comparing the composition of a number of well-studied, individual comets to that of the protostellar region IRAS 16293-2422B.  Our goal here is to determine whether other comets show similar correlations to protostellar compositions as were reported in Hale-Bopp and 67P. If such correlations are present, can comparing cometary volatile abundance ratios with
their interstellar counterparts reveal the extent of processing required to produce a comets specific composition?
How do molecular abundance ratios vary among comets in comparison to those
in protostellar regions?  Can these comparisons be used to identify
molecules that are (a) preserved from the protostellar phase, (b) 
enhanced by formation in the disk phase, or (c) depleted in comets and 
therefore destroyed in the disk phase?  Are there trends that are common to particular 
comet families?  Do particular comets or groups of comets have similar
relationships to the protostellar composition and can these relationships
be used to infer the conditions under which the various comets formed?

To address some of these questions, we compare the volatile ratios in individual comets and in the protostellar source, examining whether correlations exist between the abundances of molecules within the same molecular families (COH-, N-, and S-containing molecules). 
We then use a smaller number of comets with a common set of detected molecules to determine whether there are trends in the cometary data that indicate the relative degree to which they might have inherited interstellar material. The paper outline is as follows.
The observational data used are reviewed in Section~\ref{sec:obs}.  Only a small number of comets have sufficient observations to allow for this kind of analysis and we discuss the comets chosen in Section~\ref{sec:comet_obs}.
 Section~\ref{sec:stats} describes the statistical methods that we use. The correlation between
molecular ratios in individual comets and IRAS 16293-2422B is
examined in Section~\ref{sec:corr_family}. Using a subset of our comet database with a common set of observed molecules we report general trends in the cometary abundance ratios relative to protostellar in Section~\ref{sec:trends}.  We consider to what extent comet compositions may be a result of interstellar inheritance in Section~\ref{sec:retention} and we discuss our results in Section~\ref{sec:disc}.  Our conclusions can be found in Section~\ref{sec:conclusions}.

\section{\label{sec:obs}The observational data}

To investigate the relationship between comets and interstellar ices, we compare observations of cometary volatiles with abundances in the Class 0 protostellar binary IRAS 16293-2422B. We selected IRAS 16293-2422B over other candidate protostars for several reasons. First, its chemical richness is exceptional, with a broad inventory of volatile species, including complex organics, sulfur-bearing molecules, isotopologues, and even exotic species such as organohalogens \citep{droz19}. Second, high-resolution ALMA observations from the Protostellar Interferometric Line Survey (PILS)\citep[PILS\footnote{\url{http://youngstars.nbi.dk/PILS}}; see also][]{pils16} have provided unbiased, full spectral surveys at high angular (approximately 60 au) and spectral (approximately 0.2 km s$^{-1}$) resolution, with narrow line profiles (about 1 km s$^{-1}$) that significantly reduced line confusion and allowed precise abundance determinations \citep{jorgensen16}. Third, detailed studies comparing ROSINA measurements for comet 67P with IRAS 16293-2422 B have demonstrated that the relative abundances of CHO-, N-, and S-bearing molecules show good correlation, with some scatter, between protostellar and cometary data \citep{droz19}. Consequently, IRAS 16293-2422B provides the most robust and quantitative reference for assessing potential chemical continuity between the protostellar phase and comet populations. Here, we summarize the observational data used in this study.

\subsection{The protostellar source IRAS 16293-2422B}

IRAS 16293-2422 is a binary protostellar source, where the two protostars are separated by $\sim$ 747 au \citep{Dzib18}. The A source is more massive and more luminous (1 M$_\odot$, 18 L$_\odot$) than the B source (0.1 $M_\odot$, 3 L$_\odot$) \citep{jacobsen2018}. In IRAS 16293-2422, ices accumulated in the cold molecular cloud phase have been evaporated by the high temperatures close to the protostars. Table~\ref{tab:droz} summarizes the available observations, with column densities from D19. The majority of this data were taken using ALMA by PILS and refer to a position one beam offset in the southwest direction from source B (see D19). Infall velocity signatures are observed \citep{pineda12}, confirming that the observations probe the material falling onto the protostellar disk. 
While most of the data represent detections, there are a few molecules for which only upper limits are available. Notably, this includes NH$_3$, which is based on observations of the circumbinary envelope, and H$_2$O, which relies on an estimate based on observations of H$_2^{18}$O towards source A. For details, see Table 1 of D19.

\begin{deluxetable}{lllll}

\tabletypesize{\scriptsize}
  \tablecaption{\label{tab:droz}Molecules observed towards IRAS 16293-2422B with ALMA from the PILS survey \citep{pils16}. Column densities are taken from the summary in \cite{droz19} and references therein, with additional data from \cite{jorg20}. Binding energies are provided to indicate the relative volatilities of each species, along with their possible chemical origins. These values are primarily taken from the KIDA database\footnote{\url{https://kida.astrochem-tools.org/}}, except for (CH$_2$OH)$_2$, adopted from \cite{oberg09}, and C$_2$H$_5$SH, taken from \cite{Muller16}. Observations were made towards source B, except for water, for which the column density was estimated towards source A by \cite{persson13}. Isomers (which are not distinguishable in ROSINA observations of 67P) are indicated. The final column lists the likely formation routes of these molecules, indicating whether they are produced through gas-phase reactions, grain-surface reactions, or both. References for the grain-surface reactions are (1) \cite{garrod13}, (2) \cite{Majumdar16}, and (3) \cite{Muller16}. For gas-phase chemistry, we refer to the KIDA 2024 network \citep{kida24} for the available pathways and (4) \cite{vd14}.}
  \tablewidth{0pt}
  \tablehead{
    \colhead{ } & \colhead{Species} & \colhead{Column density } & \colhead{E$_B$} & \colhead{Possible Origins} \\
    \colhead{} & \colhead{} & \colhead{(cm$^{-2}$)} & \colhead{(K)} & \colhead{}
  }
  \startdata
  & H$_2$O & 3.3 (21) &  5600 &  Grains (hydrogenation of O)$^{[1]}$   \\
        &             &       &  & some gas phase (ion-molecule, neutral-neutral)$^{[4]}$ \\
  & CO & 1.0 (20) & 1300 & Gas-phase chemistry \\\
  & CH$_3$OH & 1.0 (19) & 5000 & Grains (successive hydrogenation of CO)$^{[1]}$   \\
  & H$_2$CO & 1.9 (18) & 4500 & Grains (hydrogenation of HCO; likely)$^{[1]}$ \\
  &         &          &      & Gas (CH$_3$ + O; plausible) \\  
  & CH$_3$CHO & 1.2 (17) & 5400 & Grains (hydrogenation of CH$_3$CO)$^{[1]}$ \\
  & HCOOH & 5.6 (16) & 5570 & Grains (HCO + OH radical recombination) \\
  \multirow{2}{*}{isomers $\begin{dcases*} \\ \\
  \end{dcases*} $}
  & C$_2$H$_5$OH & 2.3 (17) & 5400 & Grains (radical--radical CH$_3$ + CH$_2$OH)$^{[1]}$  \\
  & CH$_3$OCH$_3$ & 2.4 (17) & 3150 & Grains (radical--radical CH$_3$O + CH$_3$)$^{[1]}$ \\
  \multirow{2}{*}{isomers $\begin{dcases*} \\ \\ 
  \end{dcases*} $}
  & CH$_3$COCH$_3$ & 1.7 (16) & 3500 & Grains (radical--radical CH$_3$ + CH$_3$CO)$^{[1]}$ \\
  & C$_2$H$_5$CHO & 2.2 (15) & 4500 & Grains (radical--radical C$_2$H$_5$ + HCO)$^{[1]}$ \\
\multirow{3}{*}{isomers $\begin{dcases*} \\ \\ \\
  \end{dcases*} $}  
  & HCOOCH$_3$ & 2.6 (17) & 6295 & Grains (radical--radical CH$_3$O + HCO)$^{[1]}$ \\
  &         &          &      & Gas (O + CH$_3$OCH$_2$); Both routes: plausible \\
  & CH$_2$OHCHO & 3.2 (16) & 6295 & Grains (radical--radical CH$_2$OH + HCO)$^{[1]}$
 \\
  & CH$_3$COOH & 2.8 (15) & 6295 & Grains (CH$_3$CO + OH)$^{[1]}$ \\
  \multirow{2}{*}{isomers $\begin{dcases*} \\ \\
  \end{dcases*} $}
  & CH$_3$OCH$_2$OH & 1.4 (17) & 7500 & Grains (CH$_3$O + CH$_2$OH)$^{[1]}$  \\
  & (CH$_2$OH)$_2$ & 9.9 (16) & 7500 & Grains (CH$_2$OH + CH$_2$OH)$^{[1]}$  \\
  & NH$_2$CHO & 9.5 (15) & 6300 & Grains (NH$_2$ + HCO); Gas (NH$_2$ + H$_2$CO)$^{[1]}$ \\
  &         &          &      & Both routes: plausible \\ 
  & NH$_3$ & $<$ 6.1 (19) & 5500 & Grains (hydrogenation of N); Gas (NH$_4^+$ recombination)$^{[1]}$ \\
    &         &          &      & Both routes: plausible \\
  & HCN & 5.0 (16) & 3700 & Gas-phase chemistry \\
  & HNC & $<$ 5 (16) & 3800 & Gas-phase chemistry \\
  & CH$_3$CN & 4 (16) & 4680 & Grains (hydrogenation of C$_2$N)$^{[1]}$; Gas (CH$_3^+$ + HCN)$^{[1]}$ \\
   &         &          &      & Both routes: plausible  \\
  & HNCO & 3.7 (16) & 4400 & Grains (NH + CO)$^{[1]}$  \\
  &         &          &      & Gas (dissociative recombination of H$_2$NCO$^+$/HNCOH$^+$)  \\
  &         &          &      & Both routes: plausible  \\
  & HC$_3$N & 1.8 (14) & 4580 & Gas-phase chemistry  \\
  & H$_2$S & 1.7 (17) &  2700 & Grains (hydrogenation of S)$^{[1]}$ \\
  & OCS & 2.5 (17) & 2400 & Grains (CO + S, CS + O and SH-initiated routes)$^{[1]}$ \\
  & CS & 3.9 (15) & 3200 & Gas-phase chemistry \\
  & S$_2$ & $<$ 1.9 (16) & 2200 & Grains (energetic processing of H$_2$S-rich ices)$^{[1]}$ \\
  & SO$_2$ & 1.3 (15) & 3400 & Gas-phase chemistry  \\
  & SO & 4.4 (14) & 2800 & Gas-phase chemistry \\
  & H$_2$S$_2$ & $<$ 7.9 (14) & 3100 & Grains (radical recombination HS + HS)$^{[1]}$  \\
  & HS$_2$ & $<$ 4.4 (14) & 2650 & Grains (energetic processing of H$_2$S-rich ices)$^{[1]}$ \\
  & CH$_3$SH & 4.8 (15) & 4000 & Grains (hydrogenation of CH$_3$S)$^{[2]}$ \\
  & H$_2$CS & 1.3 (15) & 4400 & Grains (hydrogenation of CS/HCS)$^{[2]}$ \\
  & C$_2$H$_5$SH & $<$ 3.2 (15) & 6230 & Grains (C$_2$H$_5$ + SH)$^{[3]}$ \\ 
\enddata
\end{deluxetable}

\subsection{\label{sec:comet_obs}The comet sample}

Only a handful of comets have sufficient observations to allow us to make a meaningful comparison to the protostellar observations of IRAS 16293-2422B.  In addition, several molecules commonly observed in comets such as CH$_4$, C$_2$H$_6$ and C$_2$H$_2$ cannot be observed by the PILS/ALMA survey that forms our protostellar database, further limiting the number of comet observations available.    

Molecules detected in cometary coma can be split into two categories, parent or daughter. Parent molecules provide the best links to the bulk composition of the comet, and hence to the origin of the molecules in the protosolar disk or the interstellar medium.  Daughter molecules form by the thermal or photoprocessing of parent molecules,  e.g. OH, CN. Daughter molecules may also form by degradation of salts \citep[e.g. some observed NH$_3$  may have its origin in ammonium salts;][]{altwegg20}, or of macro-molecules or grains.   

Here we focus as far as possible on the abundances of parent molecules. However, for some species it is not clear whether they are parent molecules, daughter molecules or some combination of the two. For example, both direct sublimation and degradation of organic macromolecules likely contribute to the observed abundance of H$_2$CO \citep{capa15}.  Similarly, there is evidence that SO is a parent molecule, but it may also be produced by the photolysis of SO$_2$ \citep{boissier07, calmonte16}.

We include observations made at a range of wavelengths. Infrared observations are generally obtained close to the sun because comets are too faint at these wavelengths to be detected further out. The infrared allows for detection of molecules without a permanent dipole moment which are therefore undetectable in the optical or radio.  Molecules observed at these wavelengths include H$_2$O, H$_2$CO, CH$_3$OH, NH$_3$ and HCN.  At radio wavelengths molecules such as CO, H$_2$CO, CH$_3$OH, CH$_3$CN, HC$_3$N, HCN, NH$_3$, HCOOH, HNCO and H$_2$S have been observed. Neither infrared and radio observations can observe the full range of potential molecules at the same time, hence some molecules are observed in some comets but not in others.  The lack of reported detection does not necessarily mean that the molecule is not present in the comet. It could also be that it is present but not detected because of weather conditions, the wavelength range of the survey, the Doppler shift or observational time constraints. Either of these might result in a derived upper limit, although this might not necessarily be reported. The lack of a detection is therefore not necessarily an indication that the molecule is not present in a comet. 

While there is generally good consistency between abundances derived from the infrared and radio this is not the case for HCN.  This molecule shows a systematic difference of a factor of two in its mixing ratio relative to H$_2$O determined at the two wavelengths across all comets.  This discrepancy is as yet unexplained \citep{dr22}. We have chosen to use the (lower) infrared abundance ratio for those comets where both infrared and radio data are available, for consistency with those comets where HCN is only detected in the infrared.

We restrict our observational database to molecules that have been detected in more than one comet. 
We follow D19 in splitting the observed molecules into families, comparing the abundances of COH-, N- and S- bearing molecules separately. For each family we require the comets to have at least four molecules observed.  Taking this into consideration we find 12 comets for the COH-molecule  (Table~\ref{tab:dr1}) and the N-molecule (Table~\ref{tab:dr2}) families , and seven comets for the S-molecule family (Table~\ref{tab:dr3}).  The observations of all comets used here were made within 2 au of the sun, where water desorbs and therefore all other molecules observed are also desorbing at their maximum rate. Below we discuss the comets we consider in this paper.

\movetabledown=2.25in
\begin{rotatetable}
  \begin{deluxetable*}{llllllllllllllr}
    \tabletypesize{\scriptsize}
    \tablecaption{\label{tab:dr1}The
      abundances of COH-bearing species as  a percentage of
      water for the comets in our sample. \nocite{cr08,hatchell05}}
    \tablewidth{0pt}
    \tablehead{
\colhead{Comet} & \colhead{Family} & \colhead{CO} & 
 \colhead{CH$_3$OH} & \colhead{H$_2$CO} & \colhead{HCOOH} &
 \colhead{HCOOCH$_3$} & \colhead{CH$_3$CHO} & \colhead{C$_2$H$_5$OH} & \colhead{CH$_2$OHCHO} & \colhead{(CH$_2$OH)$_2$} & \colhead{C$_2$H$_5$CHO} & \colhead{H$_2$O} & \colhead{$R_h$}
}
\startdata
67P & JFC                  & 3.1$^{[1]}$ & 0.21$^{[2]}$& 0.32$^{[2]}$ & 0.013$^{[2]}$ & 0.0034\tablenotemark{a}$^{[2]}$ & 0.0047$^{[2]}$ & 0.039\tablenotemark{b}$^{[2]}$ & \nodata & 0.011\tablenotemark{d}$^{[2]}$ & 0.0047\tablenotemark{e}$^{[2]}$ & 100 & $R_h$ $<$ 1.5 au\\
73P/SW3/B & JFC & $<$1.9$^{[3]}$ & 0.54$^{[3]}$ & 0.14$^{[3]}$ & \nodata & \nodata & \nodata & \nodata & \nodata & \nodata & \nodata & 100 & $R_h$ $\sim$ 1 au\\
73P/SW3/C & JFC & 0.51$^{[3]}$ & 0.49$^{[3]}$ & 0.11$^{[3]}$ & \nodata & \nodata & \nodata & \nodata & \nodata & \nodata & \nodata & 100 & $R_h$ $\sim$ 1 au\\
C/1995 O1 (Hale-Bopp) & OCC & 23.0$^{[4]}$ & 2.4$^{[4]}$ & 1.1$^{[4]}$ & 0.09$^{[4]}$ & 0.08$^{[4]}$ & 0.025$^{[5]}$ & $<$ 0.2$^{[5]}$ & $<$ 0.04$^{[5]}$ & 0.25${[5]}$ & \nodata & 100 & $Rh$ $\sim$ 1 au\\
C/2012 F6 (Lemmon) & OCC & 4.0$^{[6]}$ & 1.6$^{[7]}$ & 0.7$^{[7]}$ & $<$ 0.07$^{[7]}$  & $<$ 0.16$^{[7]}$ & $<$ 0.07$^{[7]}$ & \nodata & $<$ 0.08$^{[7]}$ & 0.24$^{[7]}$ & \nodata & 100 & $R_h$ $\sim$ 0.8 au\\
C/2013 R1 (Lovejoy) & OCC & 7.2$^{[7]}$ & 2.6$^{[7]}$ & 0.7$^{[7]}$ & 0.12$^{[7]}$ & $<$ 0.2$^{[7]}$ & 0.1$^{[7]}$ & \nodata &  $<$ 0.07$^{[7]}$ & 0.35$^{[7]}$ & \nodata & 100 & 0.8 $<$ $R_h$ $<$ 1.1 au\\
C/2014 Q2 (Lovejoy) & OCC & 1.8$^{[8]}$  & 2.4$^{[8]}$  & 0.3$^{[8]}$ & 0.028$^{[8]}$  & 0.08$^{[8]}$  & 0.047$^{[8]}$  & 0.12$^{[8]}$  & 0.016$^{[8]}$ &  0.07$^{[8]}$ & \nodata & 100 & $R_h$ $\sim$ 1.3 au\\
153P (Ikeya-Zhang) & OCC & 5.7$^{[9]}$ & 2.5$^{[10]}$ & 0.4$^{[10]}$ & \nodata & \nodata & \nodata & \nodata & \nodata &  \nodata & \nodata & 100 & 0.51 $<$ $R_h$ $<$ 1.53 au \\
C/1996 B2 (Hyakutake) & OCC & 18.2$^{[9]}$ & 2.0$^{[11,12]}$ & 1.0$^{[11,12]}$  & \nodata & \nodata & \nodata  & \nodata & \nodata & \nodata & \nodata & 100 & 0.73 $<$ $R_h$ $<$ 1.19 au\\
C/2020 F3 (NEOWISE) & OCC & 3.2$^{[13]}$ & 2.3$^{[13]}$ & 0.3$^{[13]}$ & $<$ 0.12$^{[13]}$ & \nodata & $<$ 0.08$^{[13]}$ & \nodata & \nodata & \nodata & \nodata & 100 & 0.6 $<$ $R_h$ 1.1 au\\
C/2021 A1 (Leonard) & OCC & 1.07$^{[14]}$ & 0.88$^{[15]}$ & 0.12$^{[15]}$ & 0.19$^{[15]}$ & $<$0.15$^{[15]}$ & 0.036$^{[15]}$ & 0.18$^{[15]}$ & 0.051$^{[15]}$ & 0.13$^{[15]}$ & \nodata & 100 & 0.76 $<$ $R_h$ $<$ 1.12 au\\
C/2022 E3 (ETF) & OCC & 0.7$^{[15]}$ & 1.76$^{[15]}$ & 0.12$^{[15]}$ & 0.19$^{[15]}$ & $<$0.10$^{[15]}$ & 0.07$^{[15]}$ & 0.17$^{[15]}$  & $<$0.024$^{[15]}$ & 0.13$^{[15]}$ & \nodata & 100 & $R_h$ $\sim$ 1.1 au
\enddata
\tablenotetext{a}{Includes CH$_2$OHCHO and CH$_3$COOH} \tablenotetext{b}{Includes CH$_3$OCH$_3$} \tablenotetext{c}{Includes CH$_3$COCH$_3$} \tablenotetext{d}{Includes CH$_3$OCH$_2$OH}
\tablenotetext{e}{Includes CH$_3$COCH$_3$}
\tablerefs{[1] \cite{rubin19}, [2] \cite{schuhmann19}, [2] \cite{dr07},  [3] \cite{bm00},  [4] \cite{crovisier04},  [5] \cite{paganini14},  [6] \cite{biver14}, [7] \cite{biver15}, [8] \cite{dr16}, [9] \cite{cr08}, [10] \cite{biver99}, [11] \cite{lis97}, [12] \cite{biver22}, [13] \cite{faggi23}, [14] \cite{biver24}}
\end{deluxetable*}
\end{rotatetable}

  \begin{deluxetable*}{lllllllll}
    \tabletypesize{\scriptsize}
    \tablewidth{0pt}
    \tablecaption{\label{tab:dr2}The
      abundances of nitrogen species as  a percentage of
      water in our comet sample. \nocite{hatchell05}}
    \tablehead{
\colhead{Comet} & \colhead{Family} & \colhead{NH$_2$CHO} &
\colhead{NH$_3$} & \colhead{HCN} & \colhead{HNCO} & 
\colhead{CH$_3$CN} & \colhead{HC$_3$N} & \colhead{HNC}  \\
}
\startdata
67P & JFC & 0.004$^{[1]}$ & 0.67$^{[1]}$ & 0.14\tablenotemark{a}$^{[1]}$ & 0.027$^{[1]}$ & 0.0059$^{[1]}$ & 0.0004$^{[1]}$ & \nodata \\
73P/SW3/B & JFC & \nodata & $<$ 0.16$^{[2]}$ & 0.29$^{[2]}$ & 0.09$^{[2]}$ & 0.025$^{[2]}$ & \nodata & $<$ 0.001$^{[2]}$ \\
73P/SW3/C & JFC & \nodata & $<$ 0.30$^{[2]}$ & 0.20$^{[2]}$ & 0.08$^{[2]}$ & 0.026$^{[2]}$ & \nodata & \nodata \\
C/1995 O1 (Hale-Bopp) & OCC & 0.015$^{[3]}$  & 0.7$^{[3]}$ & 0.25$^{[3]}$ & 0.1$^{[3]}$ & 0.02$^{[3]}$ & 0.02$^{[3]}$ & 0.035$^{[3]}$ \\
C/2014 Q2 (Lovejoy) & OCC & 0.008$^{[4]}$ & 0.64$^{[5]}$ & 0.09$^{[4]}$ & 0.009$^{[4]}$ & 0.015$^{[4]}$  & 0.002$^{[4]}$  & 0.004$^{[4]}$ \\
C/2012 F6 (Lemmon) & OCC & 0.016$^{[6,7]}$ & 0.61$^{[7]}$ & 0.14$^{[6]}$ & 0.025$^{[6]}$ & \nodata & \nodata  & 0.05$^[8]$ \\
C/2013 R1 (Lovejoy) & OCC & 0.021$^{[6]}$ & 0.1$^{[7]}$ & 0.16$^{[6]}$ & 0.021$^{[6]}$ & \nodata & \nodata & 9.5 (-3)$^[9]$ \\
C/1999 B2 (Hyakutake) & OCC & \nodata & 0.5$^{[10]}$ & 0.15$^{[11]}$ & 0.07$^{[11]}$ & 0.01$^{[11]}$ & \nodata & 0.01$^{[11]}$ \\
153P (Ikeya-Zhang) & OCC & \nodata & 0.63$^{[12]}$ & 0.15$^{[13]}$ & 0.043$^{[13]}$ & 0.01$^{[13]}$ & $<$ 0.01$^{[13]}$ & 0.005$^{[13]}$  \\
C/2020 F3 (NEOWISE) & OCC & $<$ 0.033$^{[14]}$  & 0.73$^{[15]}$ & 0.12$^{[14]}$ & $<$ 0.05$^{[14]}$ & 0.011$^{[14]}$ & \nodata & 7.2 (-3)$^{[14]}$ \\
C/2021 A1 (Leonard) & OCC & 0.023$^{[16]}$ & $<$ 0.34$^{[15]}$ & 0.09$^{[16]}$ & 0.073$^{[16]}$ & 0.016$^{[16]}$ & 0.004$^{[16]}$ & 0.005$^{[16]}$ \\
C/2022 E3 (ETF) & OCC & 0.019$^{[16]}$ & \nodata & 0.09$^{[16]}$ & 0.042$^{[16]}$ & 0.017$^{[16]}$ & $<$ 0.0022$^{[16]}$ & 0.0015$^{[16]}$ 
\enddata
\tablenotetext{a}{ROSINA cannot distinguish between HCN and HNC}
\tablerefs{[1] \cite{rubin19}, [2] \cite{dr07}, [3] \cite{bm00}, [4] \cite{biver15}, [5] \cite{dr22}, [6] \cite{biver14}, [7] \cite{dr16}, [8] \cite{cordiner14}, [9] \cite{agundez14}, [10] \cite{leroy15}, [11] \cite{cr08}, [12] \cite{hatchell05}, [13] \cite{biver06}, [14] \cite{biver22}, [15] \cite{faggi23}, [16] \cite{biver24}}
\end{deluxetable*}

\movetabledown=2.25in
\begin{rotatetable}
  \begin{deluxetable*}{lllllllllllll}
    \tabletypesize{\scriptsize}
    \tablewidth{0pt}
    \tablecaption{\label{tab:dr3}The
      abundances of sulphur species as  a percentage of
      water for the comets in our sample. }
    \tablehead{
\colhead{Comet} & \colhead{Family} & \colhead{H$_2$S} &
\colhead{OCS} & \colhead{SO} & \colhead{SO$_2$} & \colhead{CS} &
\colhead{S$_2$} &\colhead{CH$_3$SH} & \colhead{H$_2$CS}  & \colhead{C$_2$H$_5$SH} & \colhead{H$_2$S$_2$} & \colhead{HS$_2$}
}
\startdata
67P & JFC & 1.1$^{[1]}$ & 0.041$^{[1]}$ & 0.071$^{[1]}$ & 0.127$^{[1]}$ & \nodata & 0.002$^{[1]}$ & 0.038$^{[1]}$ & 0.0027$^{[1]}$ & 5.8 (-4)$^{[1]}$ & $\leq$ 6.04 (-4)$^{[1]}$  & $\leq$ 1.06 (-4)$^{[1]}$ \\
C/1995 O1 (Hale-Bopp) & OCC & 1.5$^{[2]}$ & 0.4$^{[2]}$ & 0.3$^{[2]}$ & 0.23$^{[2]}$ & 0.1$^{[3]}$  & 0.02$^{[4]}$  &
\nodata & 0.02$^{[2]}$  & \nodata & \nodata & \nodata \\
C/2014 Q2 (Lovejoy) & OCC & 0.5$^{[5]}$ & 0.034$^{[5]}$ & 0.038$^{[5]}$ & \nodata & 0.043$^{[5]}$ & \nodata & \nodata & 0.013$^{[5]}$  & \nodata & \nodata & \nodata \\
C/1996 B2 (Hyakutake) & OCC & 0.8$^{[3]}$ & 0.1$^{[3]}$ & \nodata & \nodata & 0.1$^{[3]}$ & 0.01$^{[3]}$ & \nodata & \nodata &  \nodata & \nodata & \nodata \\
C/2020 F3 (NEOWISE) & OCC & 1.2$^{[6]}$ & \nodata & 0.7$^{[6]}$ & $<$ 0.29$^{[6]}$ & 0.12$^{[6]}$ & \nodata & \nodata & \nodata & \nodata & \nodata & \nodata \\
C/2021 A1 (Leonard) & OCC & 0.15$^{[7]}$ & 0.11$^{[7]}$ & $<$ 0.04$^{[7]}$ & $<$ 0.04$^{[7]}$ & 0.06$^{[7]}$ & \nodata & $<$ 0.08$^{[7]}$ & $<$ 0.03$^{[7]}$ & \nodata & \nodata & \nodata \\
C/2022 E3 (ETF) & OCC & 0.18$^{[7]}$ & 0.068$^{[7]}$ & $<$ 0.03$^{[7]}$ & $<$ 0.03$^{[7]}$ & 0.049$^{[7]}$ & \nodata & $<$ 0.06$^{[7]}$ & $<$ 0.02$^{[7]}$ &\nodata & \nodata & \nodata 
\enddata
\tablerefs{[1] Calmonte et al. (2016), [2] Bockelee-Morvan et al. (2000), [3] Charnley \& Rodgers (2008), [4] Le Roy et al. (2015), [5] Biver et al. (2015), [6] Biver et al. (2022), [7] Biver et al. (2024)}
\end{deluxetable*}
\end{rotatetable}

\newpage
\subsubsection{67P}
67P is the best-studied comet, with observations made in situ by {\it Rosetta}. We use the summary of these observations from Table 1 of D19.  The data used here were taken by the ROSINA instrument \citep{rubin19,schuhmann19,calmonte16,hadraoui19} when the comet was less than 1.5 au from the sun, and the orbiter was between 100 and 200 km of the comet surface. 

Ground-based observations of 67P are consistent with those from {\it Rosetta}, although a smaller number of molecules have been detected \citep{bonev23}.  Given this, it seems appropriate to assume that ground-based observations provide a reasonable way of determining the composition of a comet, justifying their use for the rest of our comet sample.   

\subsubsection{73P/SW3}
73P/SW3 is a short-period comet that has split into multiple components, allowing us a unique opportunity to observe separate fragments, 73P-B and 73P-C.  Infrared observations find that both fragments are poor in hydrocarbons, CO and nitrogen \citep{villanueva06,dr07,disanti07,kobayashi07}.  Radio and optical observations find very similar compositions for the two fragments \citep{crovisier09,sb11}, suggesting that the parent comet did not have significant compositional variation with depth, and hence the composition reflects primordial conditions and is not a result of chemical evolution during the comets lifetime \citep{dr07}.

The abundance ratios used here were made in the infrared \citep{dr07} when the comet was $\sim$ 1 au from the Sun.

\subsubsection{C/1995 O1 (Hale-Bopp)}
Comet C/1995 O1 (Hale-Bopp) is a dynamically old and exceptionally active comet. Infrared observations found it to be rich in hydrocarbons, CO and HCN \citep{ms99,brooke03}.  Millimeter observations were taken with IRAM by \cite{bm00} when the comet was at its most active (within 1.2 au of the sun). The values in Table~\ref{tab:dr1}--\ref{tab:dr3} for CO, CH$_3$OH,  H$_2$CO, HCOOH, HCOOCH$_3$, NH$_2$CHO, NH$_3$, HCN, HNCO, CH$_3$CN, HC$_3$N, H$_2$S, OCS, SO, SO$_2$ and H$_2$CS are from \cite{bm00}.  Millimeter observations also provide the abundance ratios of CH$_3$CHO and (CH$_2$OH)$_2$ and upper limits for C$_2$H$_5$OH and CH$_2$OHCHO \citep{crovisier04}.

\subsubsection{C/2012 F6 (Lemmon)}
C/2012 F6 (Lemmon) is a very bright, naked-eye Oort cloud comet with a high production rate of water \citep[e.g.,][]{combi14}.  The infrared observations suggest C/2012 F6 is hydrocarbon poor.  Observations with IRAM  and ALMA are in reasonable agreement with the infrared.  The data in Table~\ref{tab:dr1}--\ref{tab:dr3} are from \cite{paganini14} (CO, NH$_3$) and  \cite{biver14}.    

\subsubsection{C/2013 R1 (Lovejoy)}
C/2013 R1 (Lovejoy) is a dynamically old, long-period comet.  Optical, infrared and radio observations were taken when the comet was within 1.35 au of the sun. The optical observations of daughter molecules find this comet to be carbon-chain poor \citep{opitom15}, but infrared observations (probing the parent molecules) suggest it is hydrocarbon typical \citep{paganini14b}.  The infrared studies also show the comet to be very CO-rich, normal in HCN and CH$_3$OH, and depleted in NH$_3$ and H$_2$CO. The abundance ratios with respect to water reported in Table~\ref{tab:dr1} were made in the radio \cite{biver14} as are the detections of NH$_2$CHO and HNCO reported in Table~\ref{tab:dr2}.  The ammonia abundance is from the weighted average in \cite{dr16} (infrared) and HNC from \cite{agundez14} (radio).

\subsubsection{C/2014 Q2 (Lovejoy)}
Comet C/2014 Q2 (Lovejoy) is a long-period comet originating from the Oort cloud. It reached a perihelion of 1.29 au in Jan 2015, and was one of the most active comets since Hale-Bopp. For our purposes we use data taken when the comet was $\sim$ 1.3 au from the sun using  IRAM \citep{biver15}. The ammonia abundance is from \cite{dr22} (infrared). 

\subsubsection{153P (Ikeya-Zhang)}
Comet 153P/Ikeya-Zhang is an Oort cloud comet with an extremely long orbital period ($\sim$ 360 yrs). It has typical cometary abundances of hydrocarbons and CO \citep{dr04,disanti02,ms02,gibb03,kawakita03}. The abundances in Table~\ref{tab:dr1}  are taken from \cite{dr16} (infrared) and \cite{cr08}.  For the N-molecules the abundances were determinedin the radio with  NH$_3$ from \cite{hatchell05}, and HCN, HNCO, CH$_3$CN, HNC and HC$_3$N from \cite{biver06}.  

\subsubsection{C/1996 B2 (Hyakutake)}
C/1996 B2 Hyakutake is a dynamically young, long-period comet. Based on infrared measurements it has typical hydrocarbon and HCN abundances, and is enhanced in CO \citep{brooke96,mumma96,dr02,ms02b,disanti03}.  The abundances given in Tables~\ref{tab:dr1} -- \ref{tab:dr3}  were made in the infrared and radio and are taken from \cite{biver99}, \cite{lis97}, \cite{cr08} and \cite{dr16}.

\subsubsection{C/2020 F3 (NEOWISE)}
Comet C/2020 F3 (NEOWISE) is a very bright Oort cloud comet on a retrograde orbit, and was well studied at a variety of wavelengths. \cite{biver22} detected 8 molecules (HCN, HNC, CH$_3$OH, CS, H$_2$CO, CH$_3$CN, H$_2$ and CO) with IRAM, and \cite{faggi21} detected six parent molecules (H$_2$O, HCN, NH$_3$, CO, CH$_3$OH, H$_2$CO) in the infrared.  These abundances are summarized in Tables~\ref{tab:dr1}--\ref{tab:dr3}. 

\subsubsection{\label{sec:21A1}C/2021 A1 (Leonard)}
C/2021 A1 (Leonard) is a long-period OCC comet that reached perihelion at $R_h$ = 0.615 au. It was observed using IRAM 30m when it was between 1.22 and 0.76 au by \cite{biver24} who detected 14 species (HCN, HNC, CH$_3$CN, HNCO, NH$_2$CHO, CH$_3$OH, H$_2$CO, HCOOH, CH$_3$CHO, H$_2$S, CS, OCS, C$_2$H$_5$OH and aGg'-(CH$_2$OH)$_2$).  There were also marginal detections of HC$_3$N and CH$_2$OHCHO.  H$_2$O, CH$_3$OH, NH$_3$ and H$_2$CO were also detected with VLT/CRIRES \citep{lippi23}.  In the infrared \cite{faggi23} detected H$_2$O, CO, OCS, H$_2$CO and HCN, with upper limits for NH$_3$ and CH$_3$OH.  The upper limit derived by \cite{faggi23} for CH$_3$OH is an order of magnitude lower the abundance determined by \cite{biver24}.  The two sets of observations were taken at different times, with those of Faggi et al. being taken when $R_h$ = 0.62 au.  A potential explanation for this discrepancy was suggested by \cite{biver11} who find that there is an apparent decrease in methanol abundance as  comets get closer to the Sun arising from the depopulation of the ground vibration levels of methanol due to the increasing coma temperature.   Alternatively the cometary methanol might have all been emitted during the outburst in the period before the Faggi et al. observations were taken.

\subsubsection{C/2022 E3 (ZTF)}
\cite{biver24} also reported observations of another long-period comet, C/2022 E3 (ZTF). Using IRAM 30-m they observed the same set of 14 molecules as reported for C/2021 A1 (Section~\ref{sec:21A1}), with the observations being taken shortly after perihelion at $R_h$ = 1.11 au. In addition, they report detections of CO and H$_2$O using the ODIN 1 m space telescope.

\section{\label{sec:stats}Statistical approach}

As is often the case in astronomy, our datasets include upper limits.  These upper limits, in both the cometary and protostellar data, pose a challenge for statistical analysis, as they can introduce biases depending on how they are handled. Common approaches include excluding data points with upper limits or assigning them an arbitrary (low) value \citep[e.g.,][]{gleit85,newman89,helsel12,kafatos13}. However, both strategies can be problematic. 

{In the D19 study, the Pearson and Spearman’s rank correlation coefficients were used to assess the relationship between 67P and protostellar compositions, and regression lines were calculated using the least squares method. The dataset for comet 67P was relatively unaffected by upper limits which mainly affected only some sulphur molecules.  For the protostellar source, sulphur molecules were again affected, as was NH$_3$, a major carrier for the nitrogen in both comet and protostar. Here though, all comets in our sample have upper limits, and often for major carriers of a given element. The situation is also complicated by having upper limits data in both the cometary and protostellar data. This makes correlation analysis more complex and increases the risk of bias. 

We have approached this problem in two ways.  First, we adopted a strategy of replacing each upper limit with 50\% of its reported value.  Using this substitution, we calculated the Pearson and Spearman correlation coefficients, as well as $r^2$  and the regression line slope using the least squares method. 

Secondly, we have used statistical approaches specifically designed for datasets with upper limits, (also known as left-censored data). One such method is the Kendall correlation coefficient, which is generally preferred over Spearman’s rank for assessing the monotonic relationship between two datasets that include upper limits. For the regression analysis, we used the Akritas-Thiel-Sen (ATS) method \citep{ats} as an alternative to the least squares approach. The original Thiel-Sen method estimates the slope as the median of all pairwise slopes, excluding pairs with identical x-values. The median intercept is then determined from these pairwise slopes. The ATS method extends this by incorporating upper limits in both datasets. It determines the best-fit slope as the one that, when subtracted from the data, yields a Kendall $\tau$ value closest to zero. This approach reduces the impact of both upper limits and outliers.   We have used the implementation of the ATS method in the ``R" NADA2 library, cenken.

In the tables that follow we have included Pearson and Spearman’s rank correlation (for consistency with D19) as well as the Kendall correlation coefficients. We also report regression line slopes calculated using both the least squares and ATS methods.  For all of the statistical calculations the 1:1 points  (CH$_3$OH/CH$_3$OH, CH$_3$CN/CH$_3$CN and H$_2$S/H$_2$) are ignored, although these points are shown in the figures below. 

\section{\label{sec:corr_family}Correlations in molecular families}
D19 showed that the abundances measured in the protostellar source IRAS 16293-2422B and comet 67P are correlated. Here we expand on this prior work by comparing IRAS 16293-2422B to a number of other comets. 
To allow a direct comparison with D19 we have used the same division of the observations into molecular families.  COH-molecular abundances are given relative to the abundance of methanol, which is formed in ices by the sequential addition of hydrogen to CO \citep{hiraoka94,wk02,fuchs09}, or by reaction of CH$_3$ with OH \citep{qasim18}. It is a likely precursor to larger molecules such as glycolaldehyde and methyl formate \citep[e.g.][]{simons20,fedoseev17, chuang16,chang17}.  CH$_3$CN is used as the reference for N-molecules.  It has a similar structure to CH$_3$OH, but is likely formed in both the gas and on the grains \citep{calcutt18}. In 67P the observations cannot distinguish between CH$_3$CN and CH$_3$NC and for this source we use the total abundance of these two molecules in both the comet and the protostar. For the other comets where only CH$_3$CN has been observed we use this molecule alone as the reference.  

For sulphur, D19 used CH$_3$SH (again analogous to CH$_3$OH) as the reference molecule, but this has only been observed in 67P (with upper limits in C/2021 A1 and C/2022 E3).  We therefore need a different reference molecule.  We want to use a molecule that traces a similar region of the protostellar source as CH$_3$SH and to do this we need a molecule with a similar binding energy.  Of the S-molecules commonly observed in comets H$_2$S has a binding energy closest to that of CH$_3$SH. \cite{perrero22} calculated the binding energies of CH$_3$SH to be 4603 K on crystalline water ice, and between 2260 -- 4302 K on amorphous ice, with the values for H$_2$S being 4033 K on crystalline ice and between 1970 and 4489 K on amorphous ice.  We therefore elect to use H$_2$S as the reference molecule for the sulphur family in this work.

We follow D19 by presenting our results in log-log space because of the wide range of abundance ratios.  D19 also presented the Pearson correlation coefficient ($r$) and Spearman's rank coefficient ($\rho$) for the comparison between 67P and IRAS 16293-2422B in linear space, and found that the values were similar to the log-log comparison, with slightly higher values of $r$ for the linear comparison.   Spearman's rank $\rho$ does not depend on the scaling used.  We include the statistical parameters derived for both log-log space and for linear space in the tables that follow.

\subsection{COH-molecular abundance ratios with respect to CH$_3$OH}

We have six comets (67P, Hale-Bopp, C/2013 R1, C/2014 Q2, C/2021 A1 and C/2022 E3) in which at least seven COH-molecules have been detected.   A number of upper limits are also available for other molecules in these comets.  For C/2012 F6 there are five detections, and four upper limits. and there are  four comets (73P/SW3/C,  153P, C/2020 F3, and Hyakutake) in which we have detections of four COH-  molecules, and with an additional two upper limits for C/2020 F3. To this list we also add 73P/SW3/B in which three COH-molecules have been detected and where there is an upper limit for CO. The smaller number of observations in the latter set of comets means that any correlations need to be treated with caution.  However, we include them here because they show similar relationships to the protostellar composition as the better-studied comets. 

We follow D19 by including NH$_2$CHO in both the COH and nitrogen molecular families comparisons.  This adds an upper limit to the observations of C/2020 F3, and a detection of NH$_2$CHO to the observations of 67P, Hale-Bopp, C/2014 Q2, C/2012 F6, C/2013 R1, C/2021 A1 and C/2022 E3.  We also include HNCO in this family, adding an additional detection to all comets, except for C/2020 F3 for which there is only an upper limit. The inclusion of HNCO does not result in significant changes in the statistical parameters for 67P compared to D19, despite the latter authors not including it in their COH-molecule analysis.

Figure~\ref{fig:cho_ch3oh} compares the abundance of COH-bearing molecules relative to CH$_3$OH in comets and in IRAS 16293-2422B.  Each plot shows  the 1:1 line (dashed line) as a reference along which all points would lie if there was perfect agreement between the comet and protostellar observations. The shaded region indicates a factor of 10 on either side of this perfect agreement line. The dashed line is the best-fit regression line to the data, determined using the least-squares method, while the dotted line shows the fit using the ATS method. Each plot also gives the Pearson correlation coefficient (r), the Spearman's rank coefficient ($\rho$), and the Kendall correlation coefficient  ($\tau$).

For each of the comets shown in Figure~\ref{fig:cho_ch3oh} we can see that most of their abundance ratios fall within a factor of 10 of the protostellar value.  
An exception to this is CO in C/2014 Q2 and C/2022 E3 which are slightly more than a factor of 10 lower than  protostellar. In 67P, NH$_2$CHO, HNCO and CH$_3$CHO are enhanced by more than a factor of 10.  (CH$_2$OH)$_2$ is over-abundant in C/2012 F6 and C/2013 R1. The comet which has the largest number of molecules enhanced by more than a factor of 10 compared to IRAS 16293-2422B is C/2021 A1.  In this comet (CH$_2$OH)$_2$, HCOOCH$_3$, CH$_2$CHCHO, HNCO and NH$_2$CHO are all over-abundant.  In C/2022 E3, HCOOH and NH$_2$CHO are enhanced and CO is under-abundant. Finally, HNCO is more than a factor of 10 enhanced in both 73P/SW3/B and C.  Overall, the more complex species in  comets have higher abundances relative to CH$_3$OH than protostellar.

In general, the cometary H$_2$O/CH$_3$OH ratios are lower than seen in IRAS 16293-2422B, indicating an increase in CH$_3$OH abundances in the comets. The exceptions to this are 73P/SW3/B and C, and 67P where the cometary values are close to protostellar.  These are the three JFCs in our sample, potentially indicating less processing of CO into more complex species in this comet family, although given the small number of comets this speculation should be treated with caution.  Additionally, the  H$_2$O abundance in IRAS 16293-2422 is taken from a different position to the other molecules, lending additional uncertainty to the protostellar ratio.

We calculate the Pearson correlation, the Spearman's rank and Kendall correlation coefficients for each comet with respect to IRAS 16293-2422B. These values are given in each panel of Figure~\ref{fig:cho_ch3oh} and also in Table~\ref{tab:stats_ch3oh}. In log space all of the comets have high Pearson correlation coefficients of $>$ 0.88, indicating that the cometary and protostellar COH-molecules are related. Similarly, the Spearman's rank coefficients are also high ($\rho$ $>$ 0.79).  A Spearman's rank coefficient two-tailed significance, $p$, of $<$ 0.025 indicates that the correlation is significant at the 95\% level.  This is the case for most of the comets, with the exception of 153P and 73P/SW3/B.  The Kendall correlation coefficients are smaller than the Spearman's rank coefficients, and show good correlation at the 90\% level for all comets except 153P.

We also calculate the Pearson correlation coefficients for the linear relationship between the comet compositions and those in IRAS 16293-2422B (Table~\ref{tab:stats_ch3oh}). 
These $r$ values are all higher than seen in the log-log comparison, with $r$ $\sim$ 1 for all comets. Similar results were reported for 67P by D19.  

Figure~\ref{fig:cho_ch3oh}  shows the regression lines calculated using the least-squares  (dashed black line) and ATS (dotted black line) methods. The slopes of both lines varies among the comets, but for a given comet, the two methods generally produce similar results. Indeed, for comets Hyakutake, Hale-Bopp, C/2013 R1 and C/2014 Q2 the results are almost identical. The biggest discrepancies are the comets with the fewest observations, namely 73P/SW3/B, 73P/SW3/C and 153P.  Comet 67P has the least-squares slope that is closest to 1 (slope = 0.771). With the ATS method the highest slope is found for C/2020 F3 (with 67P the next highest). The largest deviations from a slope of 1 is for 153P (least-squares) and C/2022 E3 (ATS).

\begin{figure}
    \centering
    \includegraphics[width=0.32\linewidth,trim={0 45 0 0}]{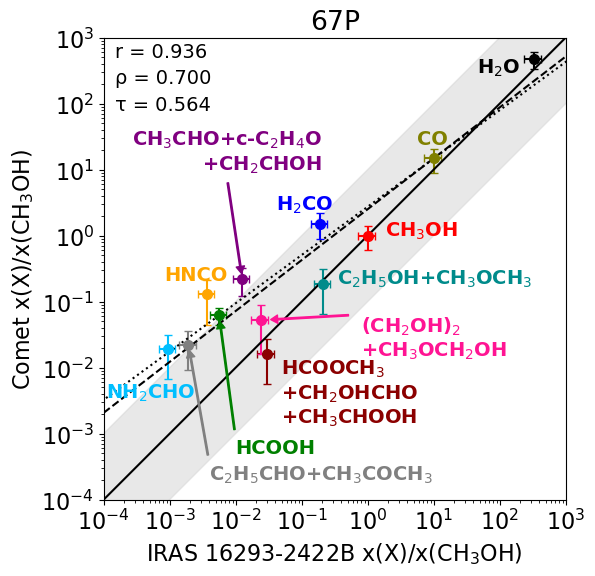}
    \hfill
    \includegraphics[width=0.32\linewidth,trim={0 45 0 0}]{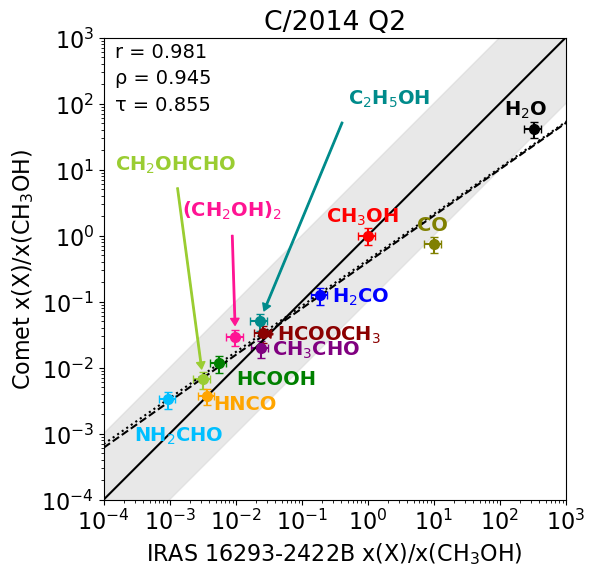} 
    \hfill    
    \includegraphics[width=0.32\linewidth,trim={0 45 0 0}]{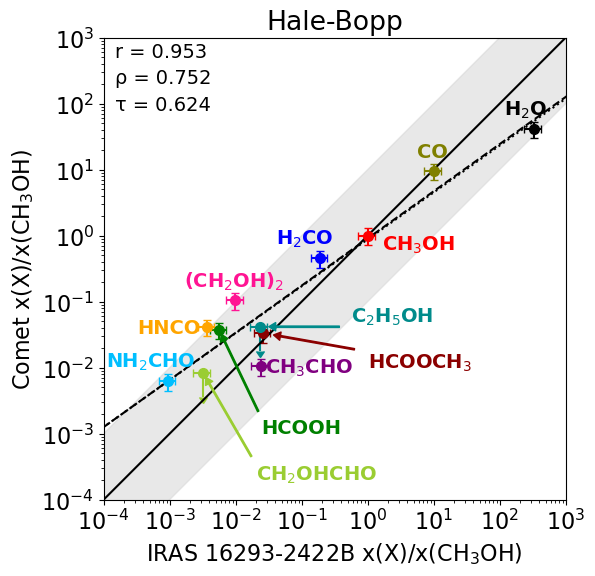}\\
    [0.8cm]
    \includegraphics[width=0.32\linewidth,trim={0 45 0 0}]{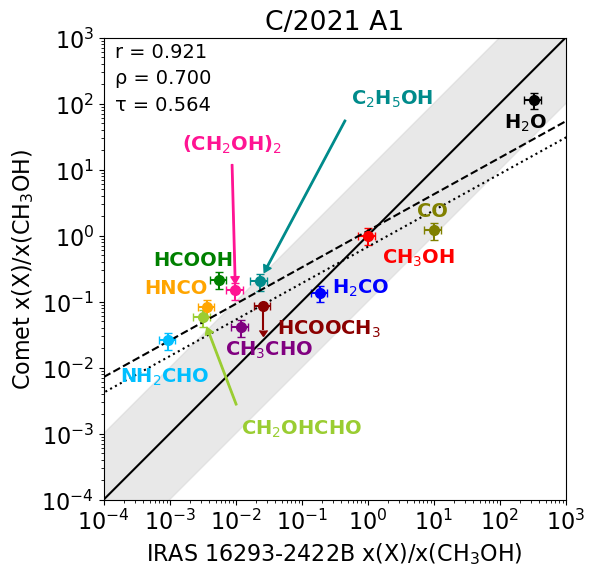}
    \hfill
    \includegraphics[width=0.32\linewidth,trim={0 45 0 0}]{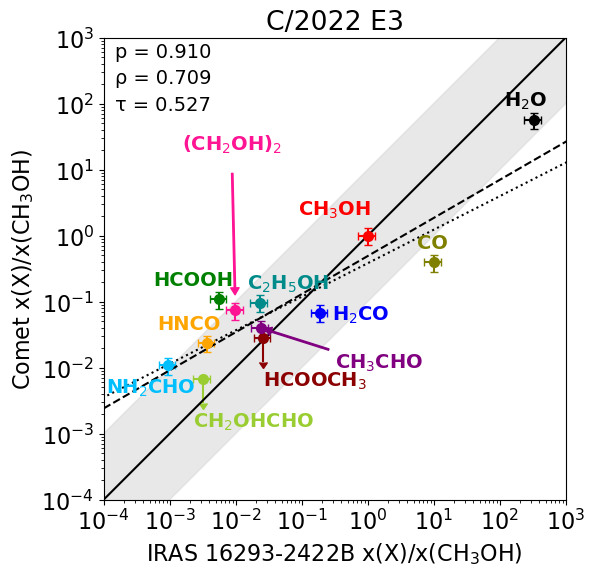}
    \hfill
    \includegraphics[width=0.32\linewidth,trim={0 45 0 0}]{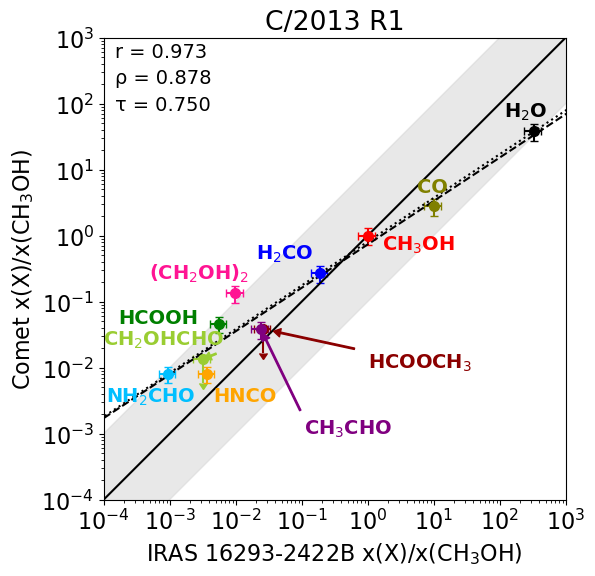}\\
    [0.8cm]
    \includegraphics[width=0.32\linewidth,trim={0 45 0 0}]{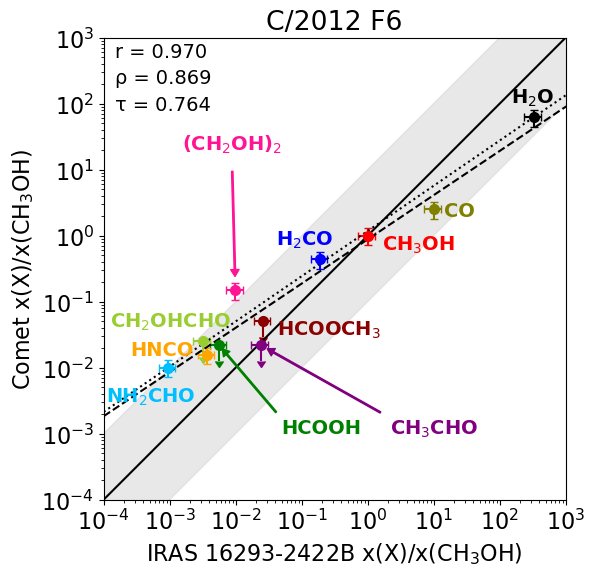}
    \hfill
    \includegraphics[width=0.32\linewidth,trim={0 45 0 0}]{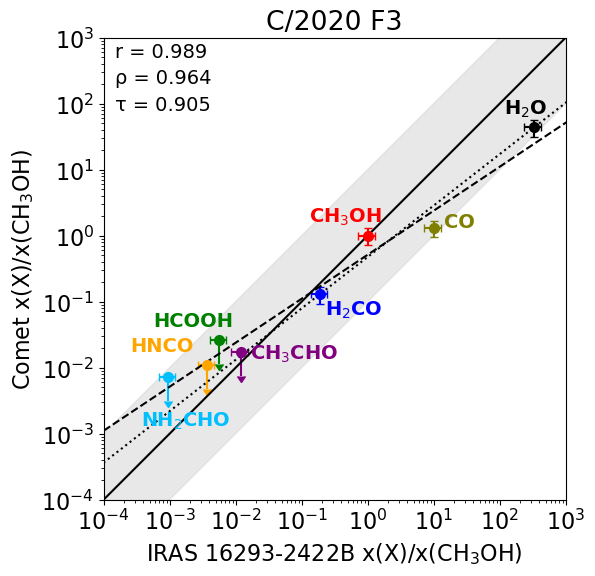}
    \hfill
    \includegraphics[width=0.32\linewidth,trim={0 45 0 0}
    ]{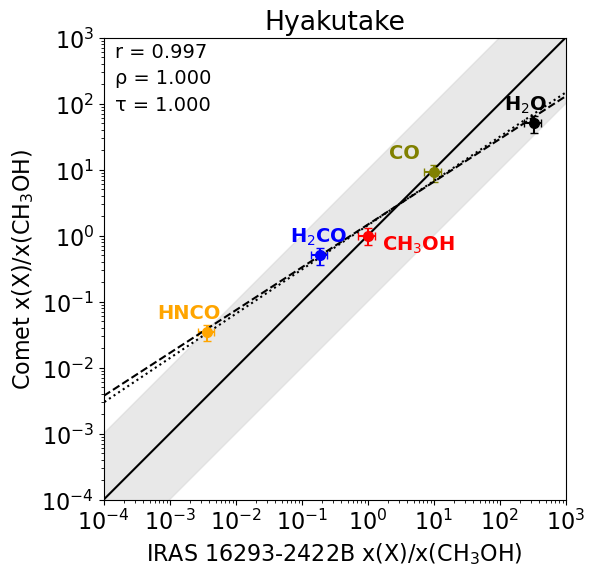}\\
    [0.8cm]
        \includegraphics[width=0.32\linewidth,trim={0 0 0 0}]{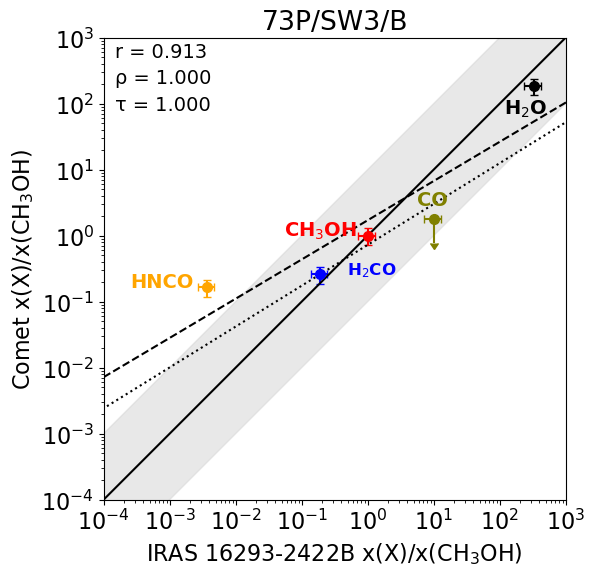}
    \hfill
    \includegraphics[width=0.32\linewidth,trim={0 0 0 0}]{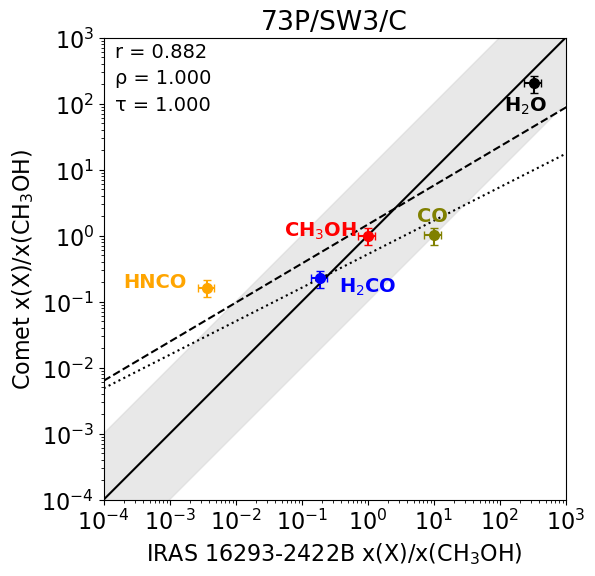}
    \hfill
        \includegraphics[width=0.32\linewidth,trim={0 0 0 0}]{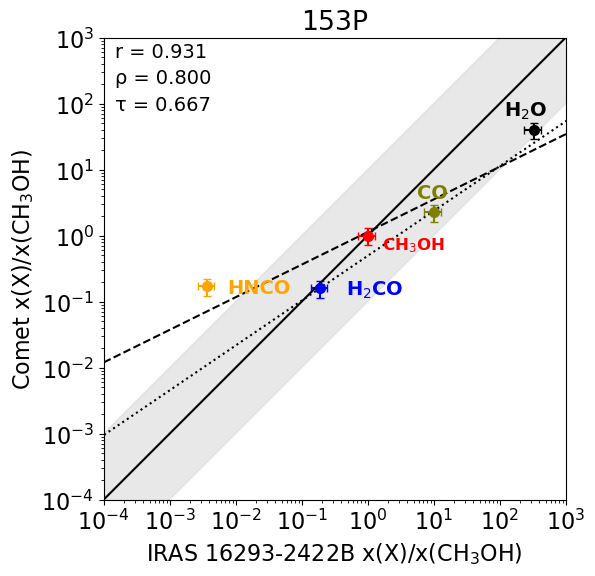}
    
    \caption{\label{fig:cho_ch3oh}Comparison of COH-molecule abundances relative to CH$_3$OH in comets and IRAS 16293-2422B.  Included in the plots are the Pearson correlation ($r$), Spearmans rank correlation ($\rho$) and Kendall correlation ($\tau$) coefficients for the log-log comparison. The solid line shows the 1:1 agreement, with the shaded area indicating a variation of a factor of 10 on either side of this. The dashed line is the regression line calculated using the least-squares method (with the upper limits replaced by 50\% of their reported value), and the dotted line is the regression line calculated using the ATS method.}
\end{figure}

\begin{deluxetable}{lcccccccccl}
    \tablecaption{\label{tab:stats_ch3oh}Statistical parameters derived from the comparison between comets and IRAS 16293-2422B for the abundance ratios of COH-abundance molecules to CH$_3$OH.  Molecules included here are those listed in Table~\ref{tab:dr1} with the addition of HNCO and NH$_2$CHO. Pearson correlation coefficients are calculated in both log and linear space.  The other statistical parameters are the same for both log and linear space. For comparison the values derived for 67P by D19 are $r$ = 0.95, $\rho$ = 0.88 in log space, and $r$ = 1.0, $\rho$ = 0.88 in linear space (D19 did not include HNCO in their calculation, but did include C$_2$H$_5$CHO and glycine which are not considered here).  The number of molecules detected in each comet are shown in the final column, with the number of upper limits in brackets.  See Section~\ref{sec:stats} for a discussion of how upper limits are treated in the statistical calculations}.
    \tablehead{
    \colhead{Comet} & \multicolumn{2}{c}{Pearson correlation} & \colhead{Spearmans} & \colhead{Spearmans} & \colhead{Kendall} & \colhead{Kendall} & \colhead{$r^2$} & \multicolumn{2}{c}{Regression} & \colhead{no. of obs.}\\
    \colhead{ } & \multicolumn{2}{c}{coefficient (r)} & \colhead{rank} & \colhead{two-tailed} & \colhead{correlation} & \colhead{two-tailed} & 
    \colhead{} & \multicolumn{2}{c}{line slope} \\
    \colhead{} & \colhead{(log) } & \colhead{(linear) } & \colhead{$\rho$} & \colhead{significance ($p$)} & \colhead{$\tau$} & \colhead{significance} & \colhead{} & \colhead{LS} & \colhead{ATS}
    }
\startdata
67P       & 0.936 & 1.000 & 0.700 & 0.016 & 0.564 & 0.017 & 0.875 & 0.771 & 0.731 & 12\\
C/2014 Q2 & 0.981 & 1.000 & 0.945 & 1.0 (-5) & 0.855 & 4.6 (-5) & 0.963 & 0.703 & 0.697 & 12\\
Hale-Bopp & 0.953 & 0.980 & 0.752 & 0.008 & 0.624 & 0.008 & 0.909 & 0.715 & 0.711 & 10 (2) \\
C/2021 A1 & 0.921 & 1.000 & 0.700 & 0.016 & 0.564 & 0.017 & 0.849 & 0.554 & 0.552 & 11 (1)\\
C/2022 E3 & 0.910 & 1.000 & 0.709 & 0.015 & 0.527 & 0.026 & 0.827 & 0.577 & 0.508 & 10 (2)\\
C/2013 R1 & 0.973 & 0.999 & 0.878 & 0.0008 & 0.750 & 0.003 & 0.947 & 0.658 & 0.663 & 9 (2) \\
C/2012 F6 & 0.970 & 1.000 & 0.869 & 0.001 & 0.764 & 0.002 & 0.940 & 0.670 & 0.685 & 7 (4)\\
C/2020 F3 & 0.989 & 1.000 & 0.964 & 0.000 & 0.905 & 0.003 & 0.978 & 0.666 & 0.779 & 4 (4)\\
Hyakutake & 0.997 & 0.989 & 1.000 & 0.000 & 1.000 & 0.083 & 0.994 & 0.649 & 0.671 & 5 \\
73P/SW3/B & 0.913 & 1.000 & 1.000 & 0.083 & 1.000 & 0.083 & 0.828 & 0.595 & 0.619 & 4 (1) \\
73P/SW3/C & 0.882 & 1.000 & 1.000 & 0.000 & 1.000 & 0.083 & 0.778 & 0.591 & 0.508 & 5 \\
153P      & 0.931 & 0.997 & 0.800 & 0.200 & 0.667 & 0.333 & 0.867 & 0.494 & 0.679 & 5 \\
\enddata
 \end{deluxetable}

\subsection{\label{sec:n}N-molecular abundance ratios relative to CH${_3}$CN}
Our sample contains 10 comets in which our reference molecule, CH$_3$CN, has been detected. Of these comets, eight have at least four detected molecules, while the remaining two have at least three detections and additional upper limits.  Figure~\ref{fig:n_ch3cn} compares the cometary and protostellar abundances and Table~\ref{tab:stats_n_ch3cn} gives the statistical parameters.

Overall while the abundance ratios of several molecules relative to CH$_3$CN fall within a factor of 10 of the protostellar values several do not. In particular, the trend is for NH$_3$/CH$_3$CN to be depleted relative to protostellar, and for HC$_3$N/CH$_3$CN to be enhanced. However, the abundance of NH$_3$ in IRAS 16293-2422B is based on an upper limit (with a value taken to be 0.5 $\times$ reported upper limit for the Pearson, Spearman's rank and $r^2$ calculations), and so it may be too high. Hence the decrease in cometary NH$_3$/CH$_3$CN compared to protostellar may be less than indicated here. HCN/CH$_3$CN also tends to be larger in comets, by $\sim$ factors of 5 -- 10. The overall pattern suggests the processing of NH$_3$ into other nitrogen-bearing molecules.

The correlation coefficients ($r$ and $\rho$) are given in Table~\ref{tab:stats_n_ch3cn}. The $r$ values are all above 0.68, while $\rho$ are $>$ 0.6, and $\tau$ $>$ 0.5. While these values are lower than for the COH-molecules, they still indicate good correlations between the cometary and interstellar observations for each comet.  The p-values for the Spearmans rank and Kendall correlations indicate the correlation is significant at $>$90\% except for Hale-Bopp.

The regression lines slopes are largely controlled by a combination of the enhancement in HC$_3$N/CH$_3$CN and the depletion of NH$_3$/CH$_3$CN in comets.  A lower NH$_3$/CH$_3$CN in the protostar would have the effect of steepening the slopes to some extent.  As with the COH-molecules the least-squares and ATS methods give similar slopes, except for C/2020 F3 where ATS gives a slope of 0.72 and the least-squares method a slope of 0.5.  This comet has the largest number of upper limits (2) and it might therefore be expected to have the greatest difference between the two methods.

\begin{figure}
    \centering
    \includegraphics[width=0.32\linewidth]{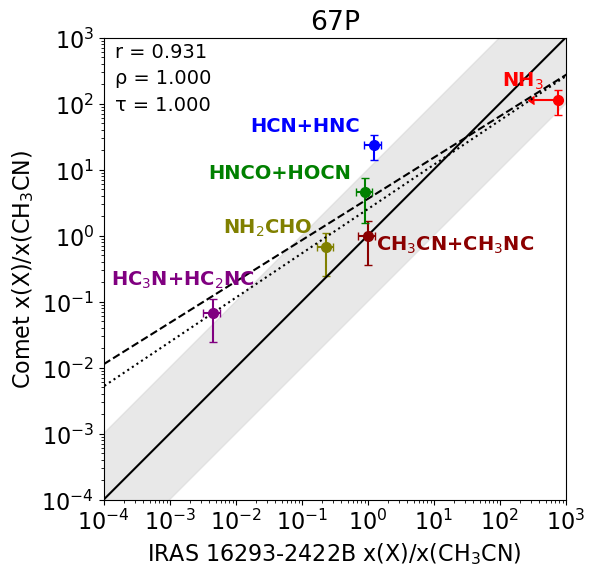}
    \hfill
    \includegraphics[width=0.32\linewidth]{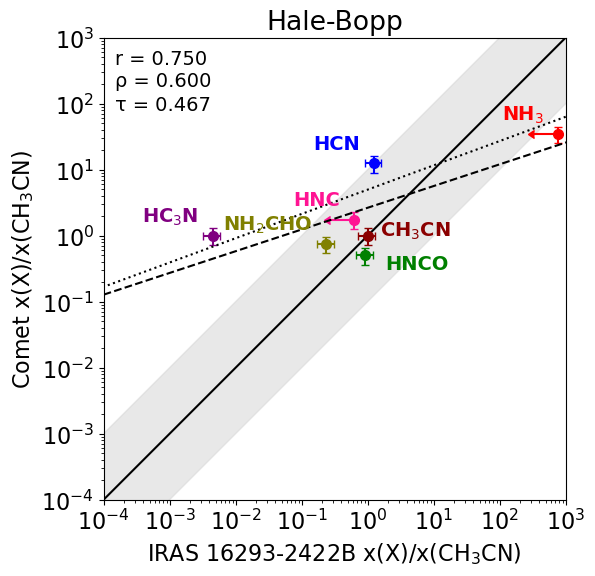}
    \hfill
    \includegraphics[width=0.32\linewidth]{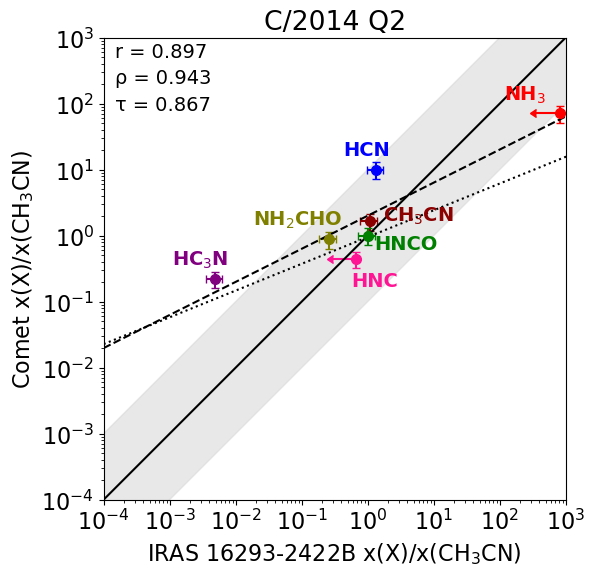}  \\
   [0.8cm] 
    \includegraphics[width=0.32\linewidth]{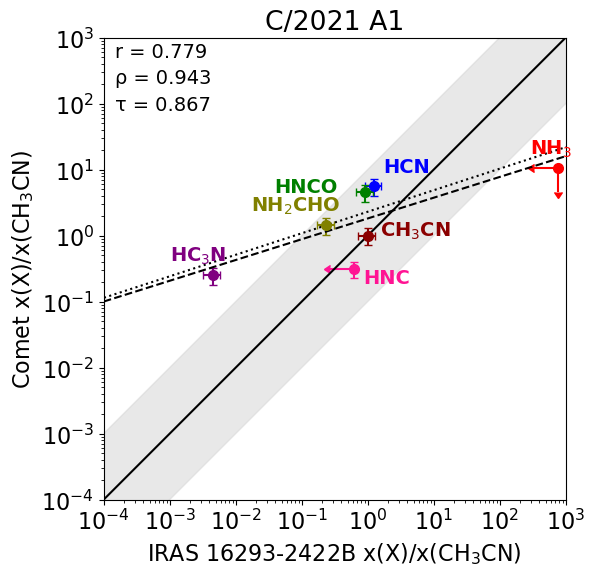} 
    \hfill
    \includegraphics[width=0.32\linewidth]{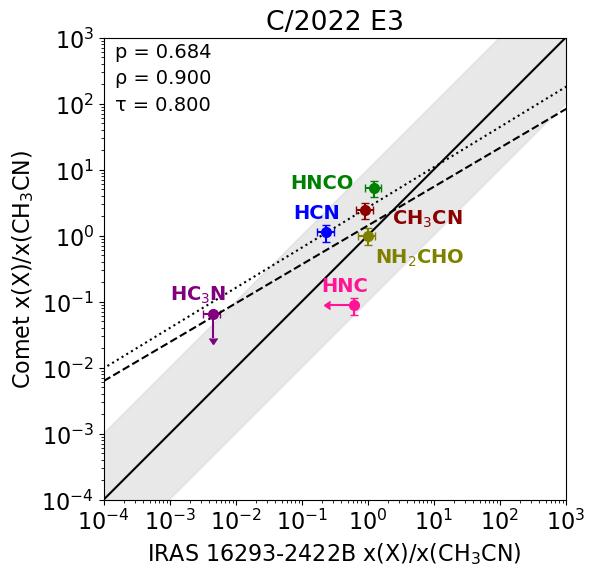}
    \hfill
    \includegraphics[width=0.32\linewidth]{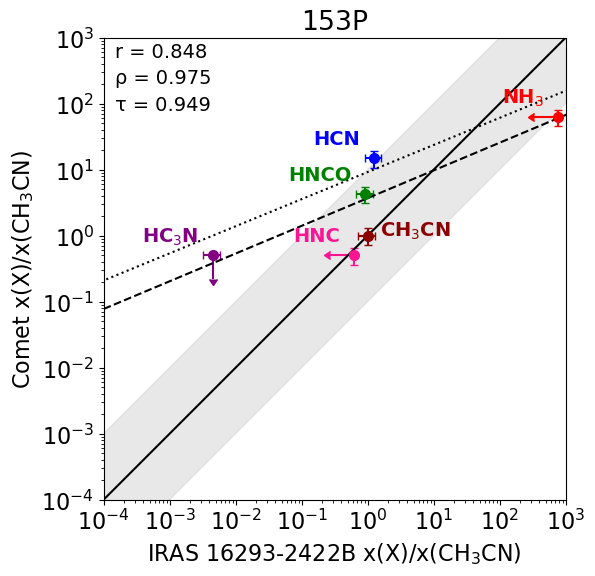}\\
    [0.8cm]
    \includegraphics[width=0.32\linewidth]{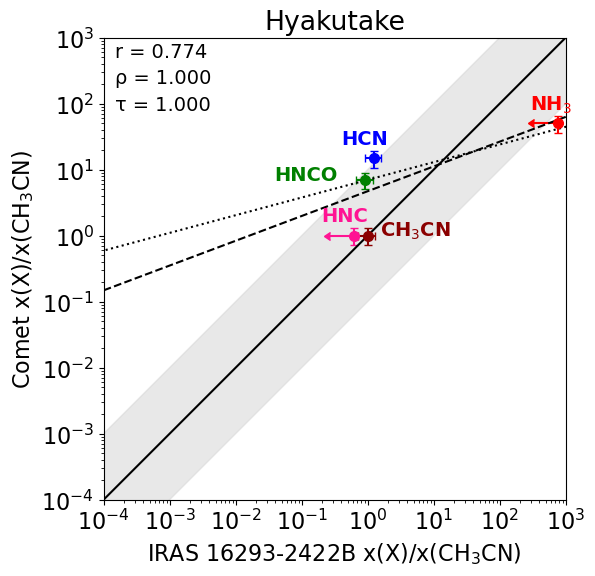} 
    \hfill
    \includegraphics[width=0.32\linewidth]{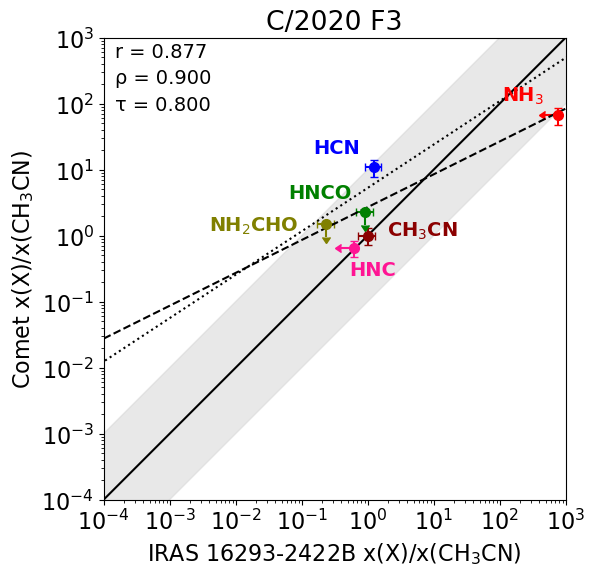}\\
 \caption{\label{fig:n_ch3cn}Comparison of N-molecule abundances in comets and IRAS 16293-2422B relative to CH$_3$CN. The best fit regression lines are shown calculated using the least-squares method (dashed line) and the ATS method (dotted line). }
\end{figure}

\begin{deluxetable}{lcccccccccl}
    \tablewidth{0pt}
    \tablecaption{\label{tab:stats_n_ch3cn}As for Table~\ref{tab:stats_ch3oh} but comparing N-abundance ratios relative to CH$_3$CN in comets and IRAS 16293-2422B.  For comparison the values derived for 67P by D19 are $r$ = 0.86 and $\rho$ = 0.93 in log space and $r$ = 0.98 and $\rho$ = 0.93 in linear space. }
    \tablehead{
    \colhead{Comet} & \multicolumn{2}{c}{Pearson correlation} & \colhead{Spearmans} & \colhead{Spearmans} & \colhead{Kendall} & \colhead{Kendall} & \colhead{$r^2$} & \multicolumn{2}{c}{Regression} & \colhead{no. of obs.}\\
    \colhead{ } & \multicolumn{2}{c}{coefficient (r)} & \colhead{rank} & \colhead{two-tailed} & \colhead{correlation} & \colhead{two-tailed} & 
    \colhead{} & \multicolumn{2}{c}{line slope} \\
    \colhead{} & \colhead{(log) } & \colhead{(linear) } & \colhead{$\rho$} & \colhead{significance ($p$)} & \colhead{$\tau$} & \colhead{significance} & \colhead{} & \colhead{LS} & \colhead{ATS}
    }
\startdata
67P       & 0.931 & 0.979 & 1.000 & 0.000 & 1.000 & 0.017 & 0.866 & 0.626 & 0.671 & 6 \\
Hale-Bopp & 0.750 & 0.943 & 0.600 & 0.200 & 0.467 & 0.272 & 0.563 & 0.329 & 0.368 & 7 \\
C/2014 Q2 & 0.897 & 0.991 & 0.943 & 0.005 & 0.867 & 0.017 & 0.805 & 0.501 & 0.405 & 7\\
C/2021 A1 & 0.779 & 0.834 & 0.943 & 0.005 & 0.867 & 0.017 & 0.607 & 0.315 & 0.326 & 6 (1)\\
C/2022 E3 & 0.684 & 0.732 & 0.900 & 0.037 & 0.800 & 0.083 & 0.479 & 0.588 & 0.609 & 5 (1) \\
153P      & 0.848 & 0.974 & 0.976 & 0.005 & 0.949 & 0.023 & 0.717 & 0.420 & 0.411 & 5 (1) \\
Hyakutake & 0.774 & 0.960 & 1.000 & 0.000 & 1.000 & 0.083 & 0.599 & 0.375 & 0.267 & 5\\
C/2020 F3 & 0.877 & 0.989 & 0.900 & 0.037 & 0.800 & 0.083 & 0.769 & 0.498 & 0.723 & 4 (2) 
\enddata
\end{deluxetable}

\subsection{S-molecular abundance ratios with respect to H$_2$S}

The correlation between cometary and protostellar S- molecules is shown in Figure~\ref{fig:s_h2s} for ratios relative to H$_2$S.   Five comets (67P, Hale-Bopp, C/2014 Q2, C/2021 A1, and C/2022 E3) have sufficient observations for correlations to be determined.  This molecular family has a larger number of upper limits, both for the comets and for IRAS 16293-2422B, adding uncertainty to any imputed correlations.  

67P  is the only comet in which CH$_3$SH was detected and we show the comparison between it and protostellar in Figure~\ref{fig:s_ch3sh}. Upper limits for CH$_3$SH are available for C/2021 A1 and C/2022 E3 but we do not consider this ratios with respect to this molecule in these comets because of the additional uncertainties this would bring. 

There is a wider scatter of abundance ratios around the regression line for the sulphur molecules than is seen for the COH- or N-molecules.  In 67P, OCS/H$_2$S, S$_2$/H$_2$S and HS$_2$/H$_2$S are under-abundant compared to protostellar, whereas SO/H$_2$S and SO$_2$/H$_2$S are enhanced by more than a factor of 10.  SO$_2$/H$_2$S and SO/H$_2$S are also enhanced in Hale-Bopp, with the increase in SO/H$_2$S being approximately a factor of 100.  In C/2014 Q2, OCS/H$_2$S is more than a factor of 10 lower than in IRAS 16293-2422B, whereas SO/H$_2$S is much higher.  In Hyakutake OCS/H$_2$S is slightly under-abundant compared to protostellar, but the other three molecules observed in this comet are in good agreement.  In general the trend seems to be that OCS/H$_2$S  and S$_2$/H$_2$S are reduced in comets, and SO/H$_2$S and SO$_2$/H$_2$S are enhanced.

The larger variation in abundance ratios compared to protostellar is reflected in the values of the Pearson and Spearmans rank coefficients, which are lower than for the other molecular families.  Hale-Bopp has the lowest value of $r$ at 0.19, whereas 67P,  C/2021 A1, C/2022 E3, and C/2014 Q2 have $r$ of 0.44, 0.92, 0.43, and 0.21 respectively.  The Spearman's rank and Kendall correlation coefficients for Hale-Bopp and C/2014 Q2 are either zero or close to it, indicating  no correlation with IRAS 16293-2422B. Some degree of correlation can be seen for 67P and C/2022 E3 where $\rho$ $>$ 0.37, and $\tau$ $>$ 0.31, but the p-values for both correlation coefficients are high. Similarly weak correlation is seen when considering abundance ratios in 67P relative to CH$_3$SH.  The best correlation is seen for C/2021 A1 with $p$ values for both Spearman's rank and Kendall correlations indicating correlations significant at the 90\% level.  

\begin{figure}
   \centering
    \includegraphics[width=0.3\linewidth]{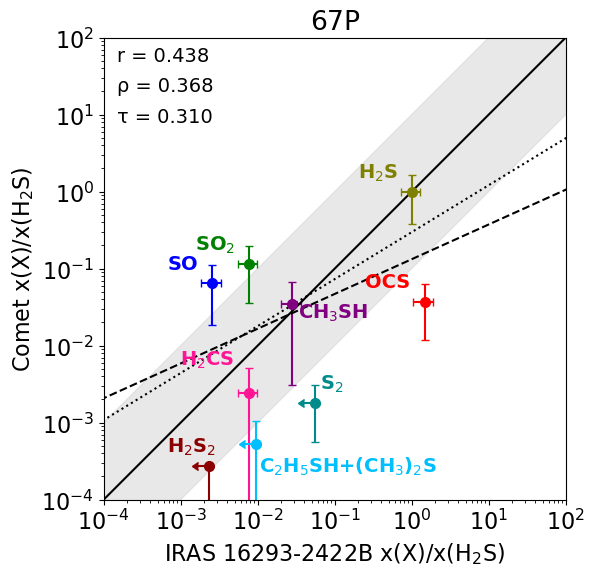}
    \hfill
    \includegraphics[width=0.3\linewidth]{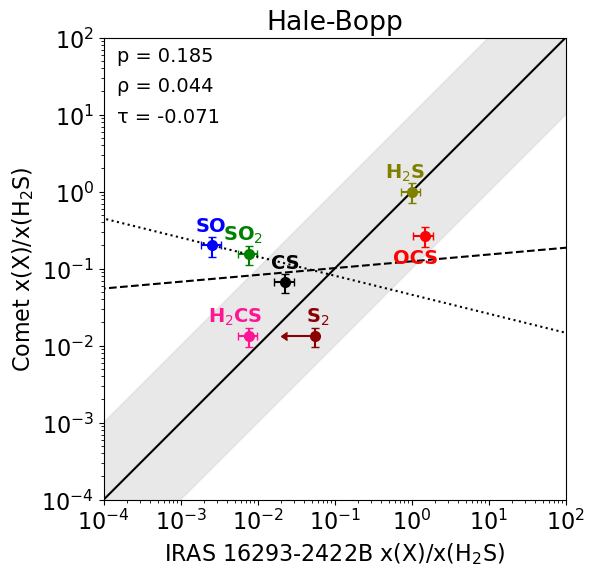}
    \hfill
       \includegraphics[width=0.3\linewidth]{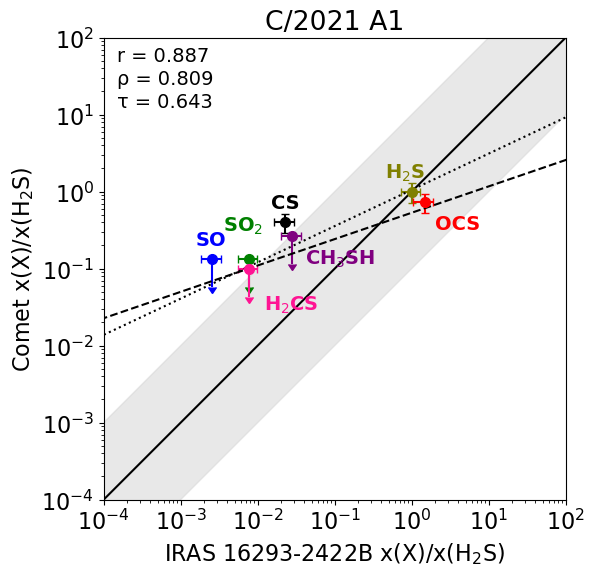}\\
        [0.8cm]
        \includegraphics[width=0.3\linewidth]{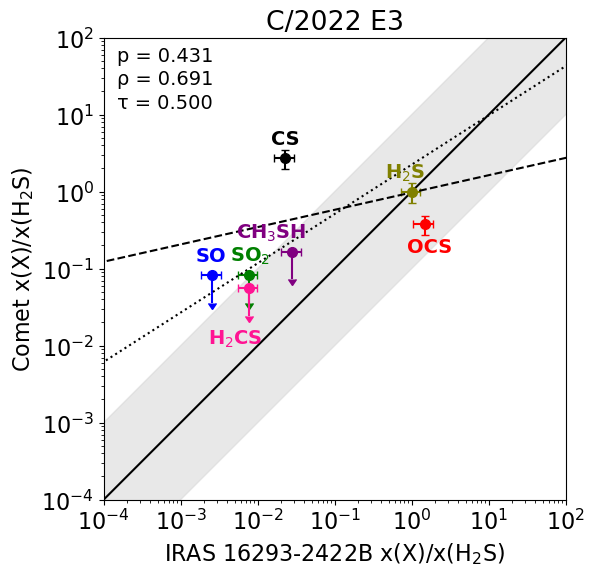}
    \hfill 
    \includegraphics[width=0.3\linewidth]{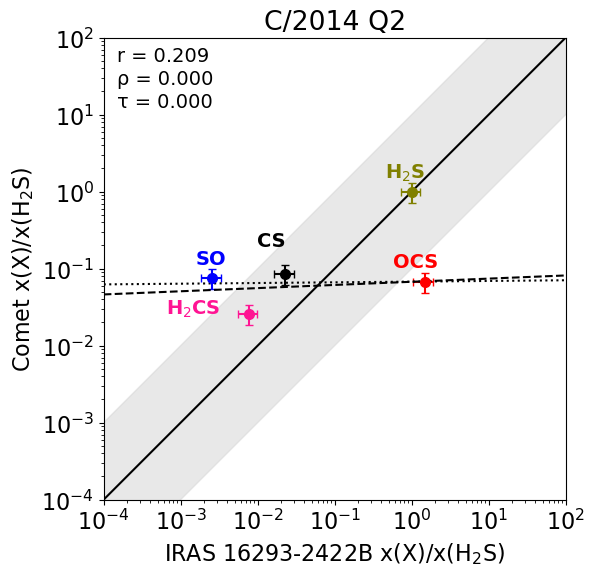}
    \hfill
    \caption{\label{fig:s_h2s}Comparison of S-molecule abundances in comets and IRAS 16293-2422B relative to H$_2$S. The best fit regression lines are shown calculated using the least-squares method (dashed line) and the ATS method (dotted line). }
\end{figure}

\begin{figure}
    \includegraphics[width=0.3\linewidth]{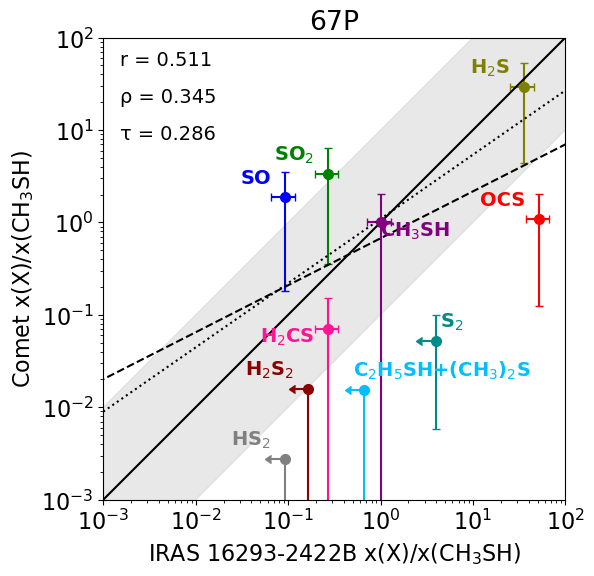}
    \caption{\label{fig:s_ch3sh}Comparison of S-molecule abundances relative to CH$_3$SH in 67P with those in IRAS 16293-2422B. The best fit regression lines are shown calculated using the least-squares method (dashed line) and the ATS method (dotted line). }
\end{figure}

\begin{deluxetable}{lcccccccccl}
    \tablewidth{0pt}
    \tablecaption{\label{tab:stats_s_h2s}As for Table~\ref{tab:stats_ch3oh} but showing statistical parameters derived from the comparison between comets and IRAS 16293-2422B for S-abundance ratios relative to H$_2$S.  For comparison, the values derived for 67P (relative to CH$_3$SH) by D19 are $r$ = 0.49, $\rho$ = 0.32 in log space, and $r$ = 0.5, $\rho$ = 0.32 in linear space.  }
    \tablehead{
    \colhead{Comet} & \multicolumn{2}{c}{Pearson correlation} & \colhead{Spearmans} & \colhead{Spearmans} & \colhead{Kendall} & \colhead{Kendall} & \colhead{$r^2$} & \multicolumn{2}{c}{Regression} & \colhead{no. of obs.}\\
    \colhead{ } & \multicolumn{2}{c}{coefficient (r)} & \colhead{rank} & \colhead{two-tailed} & \colhead{correlation} & \colhead{two-tailed} & 
    \colhead{} & \multicolumn{2}{c}{line slope} \\
    \colhead{} & \colhead{(log) } & \colhead{(linear) } & \colhead{$\rho$} & \colhead{significance ($p$)} & \colhead{$\tau$} & \colhead{significance} & \colhead{} & \colhead{LS} & \colhead{ATS}
    }
\startdata
67P (rel. to H$_2$S) & 0.438 & 0.499 & 0.368 & 0.330 & 0.310 & 0.249 & 0.191 & 0.564 & 0.518 & 8 (2) \\
67P (rel. to CH$_3$SH) & 0.511 & 0.488 & 0.345 & 0.364 & 0.286 & 0.292 & 0.261 & 0.635 & 0.526 & 9 (2) \\
Hale-Bopp & 0.185 & 0.607 & 0.044 & 0.934 & -0.071 & 0.845 &  0.034 & 0.110 & -0.247 & 7\\
C/2014 Q2 & 0.209 & 0.409 & 0.000 & 1.000 & 0.000 & 1.000 & 0.043 & 0.0411 & 0.009 & 5\\
C/2021 A1  & 0.917 & 0.826 & 0.845 & 0.051 & 0.650 & 0.080 & 0.762 & 0.292 & 0.471 & 3 (4)\\
C/2022 E3  & 0.431 & -0.001 & 0.691 & 0.128 & 0.500 & 0.173 & 0.185 & 0.281 & 0.640 & 3 (4)
\enddata
\end{deluxetable}

\section{\label{sec:trends}Trends in abundance ratio variations between comets and IRAS 16293-2422B}

As can be seen above, comets show a wide range of compositions, and not all molecules are detected in every comet.  The inclusion or exclusion of particular molecules can affect the derived statistical parameters, and influence any trends we derive from a comparison of the cometary and protostellar abundance ratios. To address this we now consider a subset of comets with a common set of molecular detections (Table~\ref{tab:common}). Again we split the observations into molecular families and compare abundances to those observed in IRAS 16293-2422B.  We take two approaches. First, in this section we make a qualitative comparison of cometary compositions to IRAS 16293-2422B and then in  Section~\ref{sec:retention} we consider if there are any quantitative assessments that can be made to estimate the relative degree to which cometary material is inherited.

\begin{deluxetable}{lll}
    \tablewidth{0pt}
    \tablecaption{\label{tab:common}The subset of comets with a common set of detected molecules}
    \tablehead{
    \colhead{Family} & \colhead{Comets} & \colhead{Molecules}
    }
    \startdata
    COH-molecules & 67P, Hale-Bopp, & CO, CH$_3$OH, H$_2$CO, HCOOH,\\
    & C/2013 R1, C/2014 Q2, & CH$_3$CHO, (CH$_2$OH)$_2$, \\
    & C/2021 A1, C/2022 E3 & NH$_2$CHO, HNCO \\
    \\
    N-molecules & 67P, Hale-Bopp, & NH$_2$CHO, HNCO, HCN,\\
    & C/2014 Q2, C/2021 A1 & CH$_3$CN, H$_3$CN \\
    \\
    S-molecules & 67P, Hale-Bopp & H$_2$S, OCS, SO, H$_2$CS \\
    & C/2014 Q2 & 
    \enddata
\end{deluxetable}
\subsection{COH-molecules}
 For the COH-molecular family we choose the comets 67P, Hale-Bopp, C/2013 R1, C/2014 Q2, C/2021 A1 and C/2022 E3.  Each of these comets has detections of CO, CH$_3$OH, H$_2$CO, HCOOCH, CH$_3$OH, (CH$_2$OH)$_2$, HNCO, NH$_2$CHO and H$_2$O.  
Here we consider how the carbon and oxygen is partitioned between these molecules in each comet and in IRAS 16293-2422B.  We exclude H$_2$O because it is by far the most abundant molecule, making up $>$ 95\% of the COH molecules in all of the comets ($>$ 75\% in Hale-Bopp) and would therefore dominate the comparisons.  In addition, the H$_2$O observation in IRAS 16293-2422B is from source A, rather than from source B which is used for the other molecular detections.  With the other eight molecules we determine their total abundance relative to H$_2$O, and then construct pie charts (Figure~\ref{fig:common_coh_pie}) that show the relative contribution of each molecule to this total carbon and oxygen budget in the set of eight common molecules.  

\begin{figure}
    \includegraphics[width=0.24\linewidth,trim=2.5cm 1.5cm 0.8cm 0cm,clip]{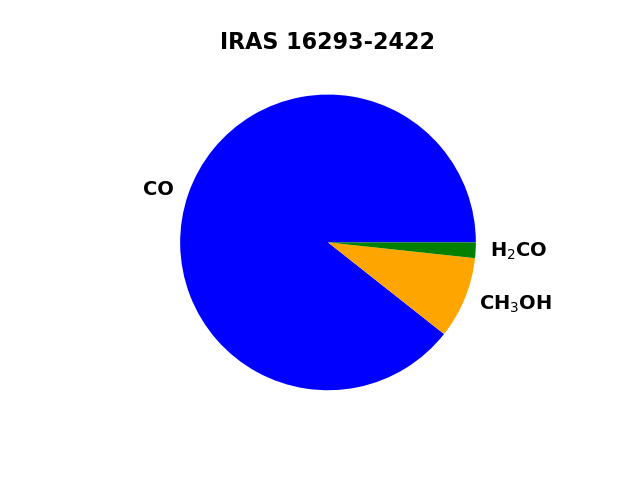}
    \\[1cm]
    \includegraphics[width=0.24\linewidth,trim=2.5cm 1.5cm 0.8cm 0cm,clip]{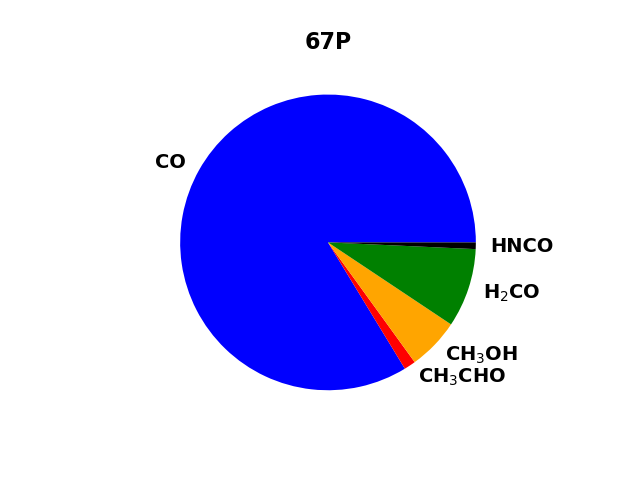}
    \includegraphics[width=0.24\linewidth,trim=2.5cm 1.5cm 0.8cm 0cm,clip]{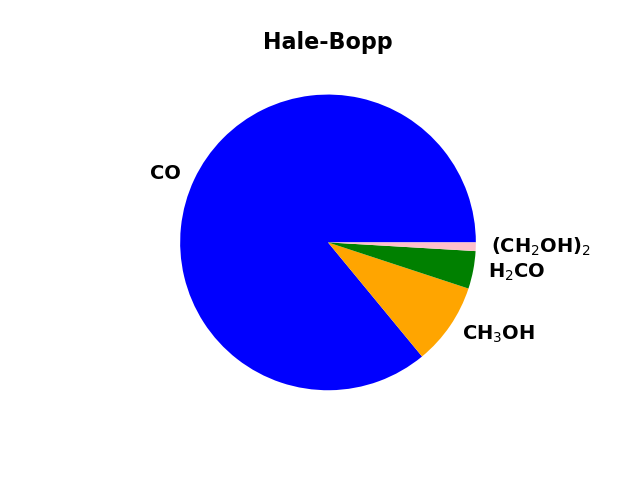}
    \includegraphics[width=0.24\linewidth,trim=2.5cm 1.5cm 0.8cm 0cm,clip]{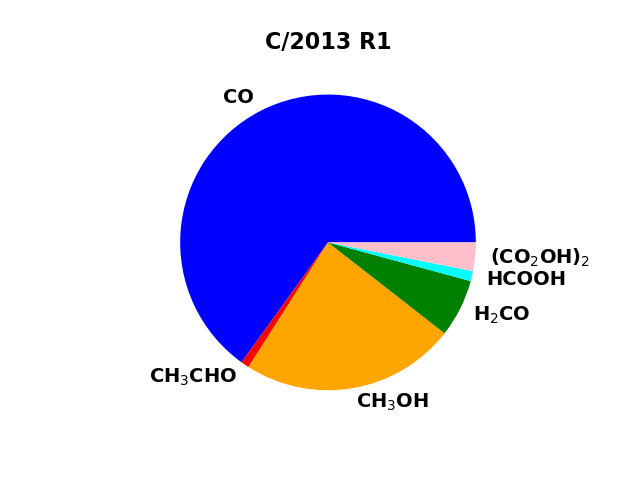}
    \\[1cm]
    \includegraphics[width=0.24\linewidth,trim=2.5cm 1.5cm 0.8cm 0cm,clip]{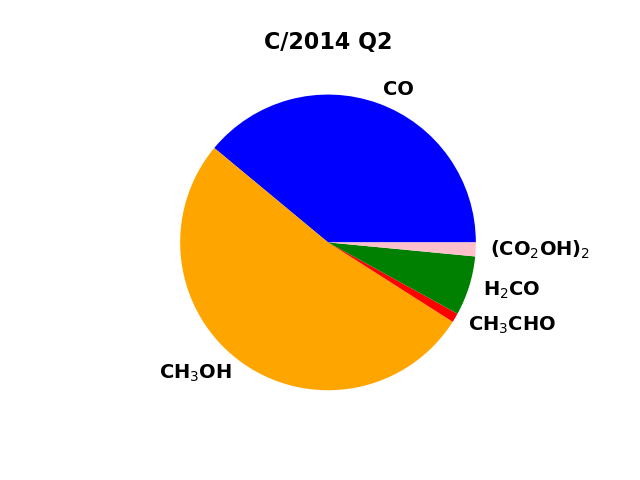}
    \includegraphics[width=0.24\linewidth,trim=2.5cm 1.5cm 0.8cm 0cm,clip]{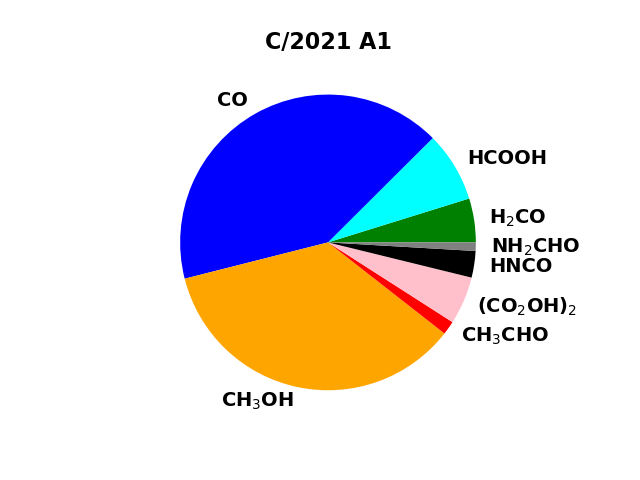}
    \includegraphics[width=0.24\linewidth,trim=2.5cm 1.5cm 0.8cm 0cm,clip]{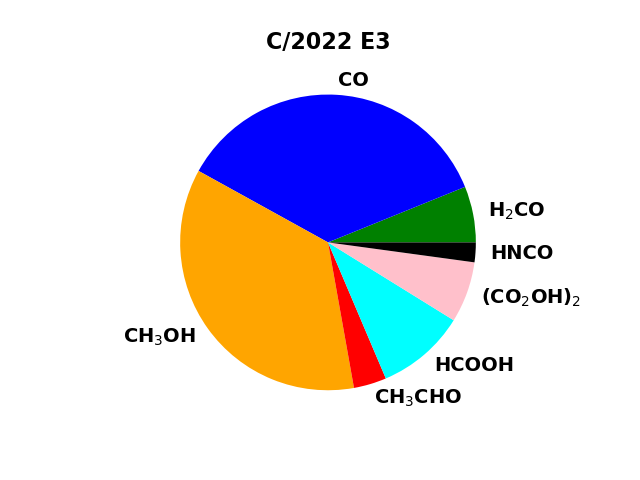}
    
    \caption{\label{fig:common_coh_pie}Division of carbon and oxygen among the molecules CO, H$_2$CO, CH$_3$OH, HCOOH, CH$_3$CHO, (CH$_2$OH)$_2$, HNCO and NH$_2$CHO.}
\end{figure}
It should be noted that the molecules discussed here do not represent the entirety of the carbon-oxygen budget in comets (or in the interstellar medium). In particular, common cometary molecules such as CO$_2$, C$_2$H$_2$ and C$_2$H$_6$ are not included because they cannot be detected by the PILS survey that provides the observations of IRAS 16293-2422B used here. These pie charts therefore do not represent the distribution of the total oxygen and carbon budget among all possible molecules.

In IRAS 16293-2422B the composition is dominated by CO (89\%), H$_2$CO (1.7\%) and CH$_3$OH (8.9\%), with small abundances of more complex molecules. The comets show a wide range of compositions.  67P and Hale-Bopp have compositions that are closest to IRAS 16293-2422B, with CO fractions of 83\% and 85\% respectively. The fraction of CH$_3$OH is protostellar in Hale-Bopp, but lower in 67P.  Both comets have more H$_2$CO than protostellar.  In the other comets CO is lower than protostellar and CH$_3$OH is larger. Indeed, CH$_3$OH is the dominant CHO-molecule in C/2014 Q2, and has a similar abundance to CO in C/2021 A1 and C/2022 E3.

All of the comets contain more complex molecules than the protostar, with COMs making up $\sim$ 25\% of C/2021 A1 and C/2022 E3.  The increase in cometary abundances of COMs is accompanied by a decrease in the abundance of CO, suggesting that CO has been processed into the more complex organics by grain chemistry in the disk. For example, CO can be sequentially hydrogenated to form H$_2$CO and CH$_3$OH, which can subsequently be transformed into larger COMs through thermal processing or irradiation by cosmic rays or UV photons.  A similar pattern of reduced CO and enhanced CH$_3$OH and H$_2$CO is seen in the other comets in Table~\ref{tab:dr1}, possibly suggesting that complex species might exist in these comets too, even though these molecules were not detected at the time of their apparition.

\subsection{N-molecules}
For the N-molecule family we have four comets (67P, Hale-Bopp, C/2014 Q2, and C/2021 A1) in which NH$_2$CHO, HCN, HNCO, CH$_3$CN and HC$_3$N are observed.   We do not include the common cometary molecule NH$_3$  because only an upper limit is available in IRAS 16293-2422B.  The abundances of these molecules as a fraction of total N abundance contained in NH$_2$CHO, HCN, HNCO, CH$_3$CN and HC$_3$N are shown in Figure~\ref{fig:common_n_pie}.

\begin{figure}
    \includegraphics[width=0.24\linewidth,trim=2.5cm 1.5cm 0.8cm 0cm,clip]{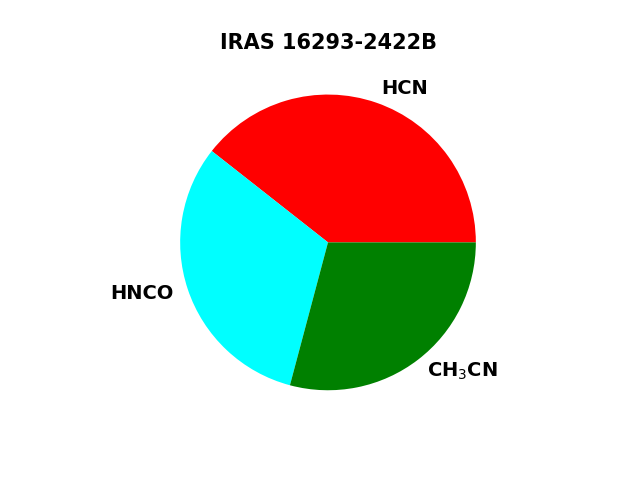}
    \\[1cm]
    \includegraphics[width=0.24\linewidth,trim=2.5cm 1.5cm 0.8cm 0cm,clip]{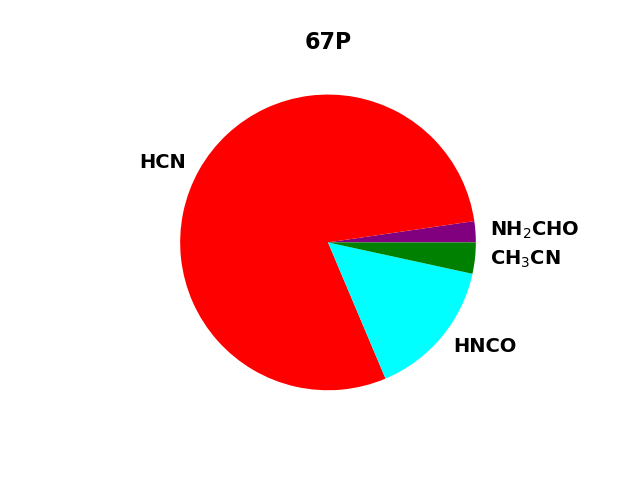}
    \includegraphics[width=0.24\linewidth,trim=2.5cm 1.5cm 0.8cm 0cm,clip]{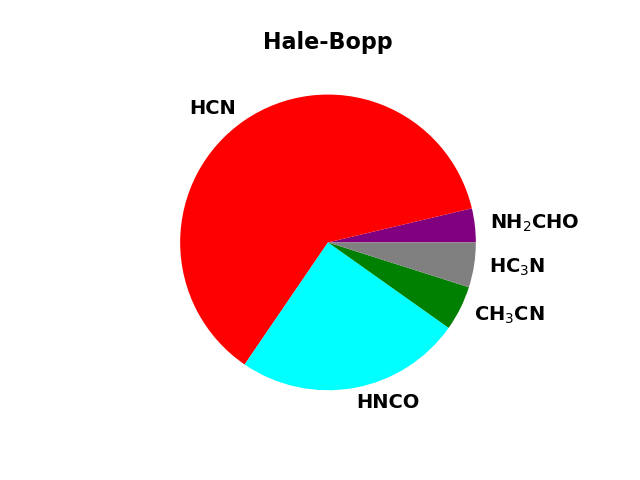}
    \includegraphics[width=0.24\linewidth,trim=2.5cm 1.5cm 0.8cm 0cm,clip]{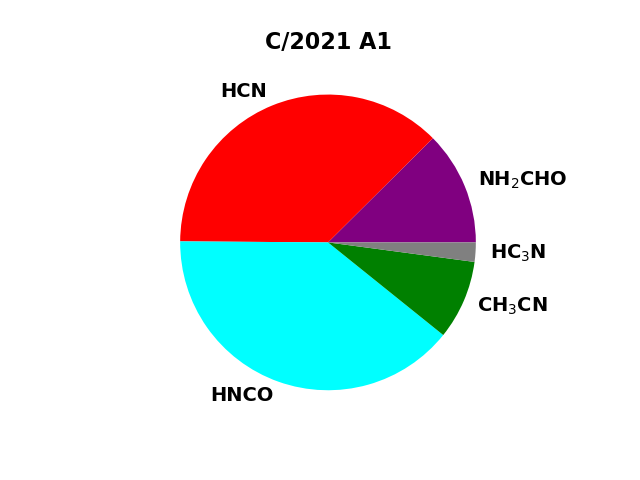}
    \includegraphics[width=0.24\linewidth,trim=2.5cm 1.5cm 0.8cm 0cm,clip]{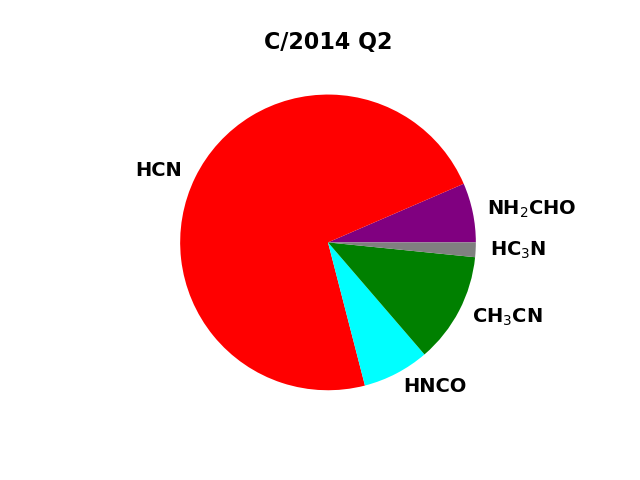}
    \caption{\label{fig:common_n_pie}Division of nitrogen among the molecules HCN, NH$_2$CHO, HNCO, CH$_3$CN, and HC$_3$N.}
\end{figure}

For our common set of observed molecules the nitrogen is roughly evenly split between HNCO, HCN and CH$_3$CN in the protostellar source.  In comets  CH$_3$CN is greatly reduced, and most show an increase in the abundance of HCN  and a reduction in HNCO.  The situation is a little different in C/2021 A1 where HCN and HNCO are both slightly enhanced over protostellar, and CH$_3$CN is reduced, with an increase in NH$_2$CHO and HC$_3$N. All of the comets  have higher abundances of NH$_2$CHO.  Like the COH-molecules the N-molecule data suggests disk processing, to varying degrees, is required to account for the cometary compositions.

One way this could happen is by photolysis of inherited ices in the disk. Experiments suggest that photolysis of nitrogen-bearing molecules in water ice can lead to the formation of other molecules such as imines, amines and amides, e.g. \cite{ligerink17,ligterink18,jones11,raunier04,islam14,fedoseev16}. 
In particular, \cite{bulak21} irradiated CH$_3$CN in water and produced HCN, HNCO and NH$_2$CHO, as well as larger N-bearing COMs.  

\subsection{S-molecules}

For sulphur molecules we have three comets in which H$_2$S, SO, H$_2$CS and OCS have been detected.  Of these four, OCS and H$_2$S are the major sulphur-bearing molecules found in IRAS 16293-2422B.  H$_2$S is also the dominant molecule in comets, but the abundance of OCS is lower. In addition, the abundances H$_2$CS and SO are higher in comets. 

\begin{figure}
    \includegraphics[width=0.24\linewidth,trim=3cm 1.5cm 0.8cm 0cm,clip]{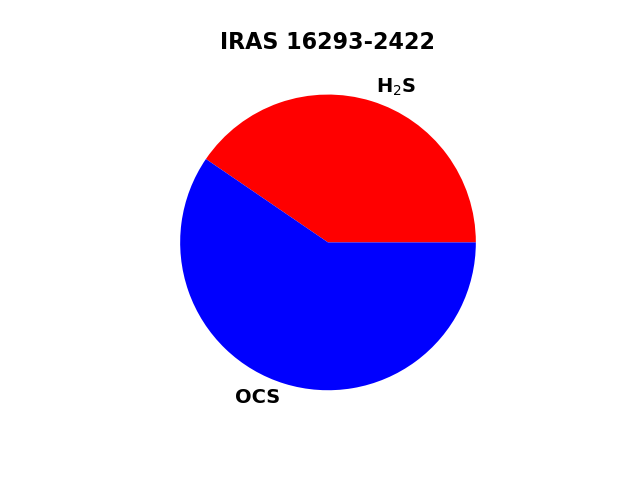}
    \\[1cm]
    \includegraphics[width=0.24\linewidth,trim=2.5cm 1.5cm 0.8cm 0cm,clip]{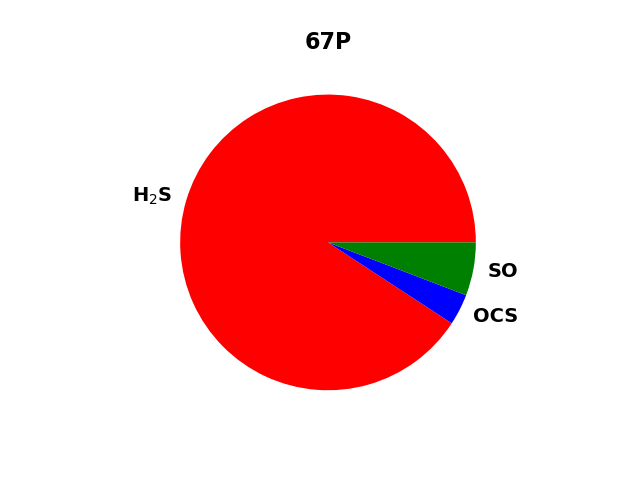}
    \includegraphics[width=0.24\linewidth,trim=2.5cm 1.5cm 0.8cm 0cm,clip]{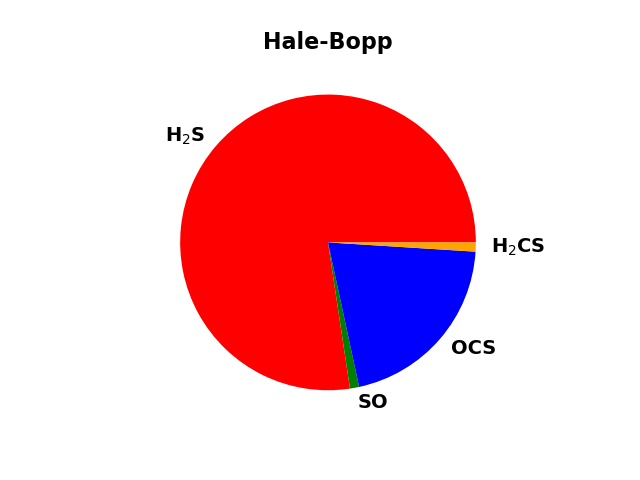}
    \includegraphics[width=0.24\linewidth,trim=2.5cm 1.5cm 0.8cm 0cm,clip]{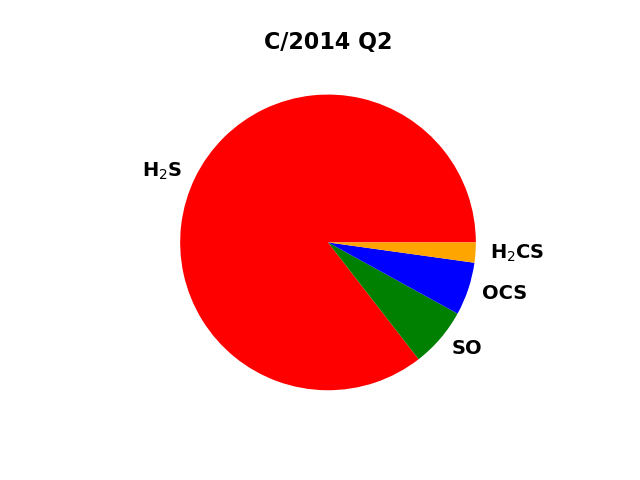}
    \\[1cm]
    \includegraphics[width=0.24\linewidth,trim=2.5cm 1.5cm 0.8cm 0cm,clip]{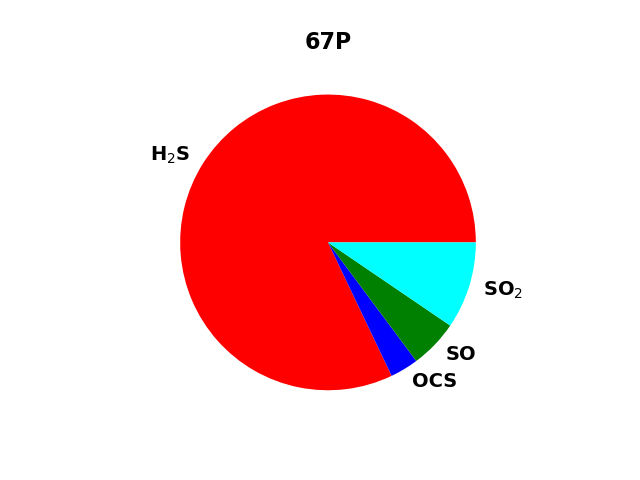}
    \includegraphics[width=0.24\linewidth,trim=2.5cm 1.5cm 0.8cm 0cm,clip]{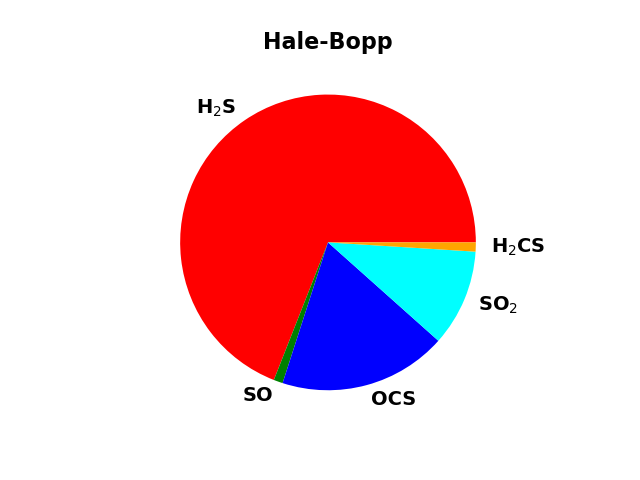}    \\ [1cm]
    \caption{\label{fig:common_s_pie}{\it top} Major carriers of sulphur in IRAS 16293-2422B.  {\it middle} Division of sulphur among the molecules H$_2$S, SO, H$_2$CS and OCS. {\it bottom} Division of sulphur among molecules H$_2$S, SO, SO$_2$, H$_2$CS and OCS. }
\end{figure}

IRAS 16293-2422B and two comets (67P and Hale-Bopp) also contain SO$_2$.  The partition of sulphur between molecules including SO$_2$ is shown in Figure~\ref{fig:common_s_pie} (bottom line).  There is no change to the distribution of sulphur in IRAS 16293-2422B where SO$_2$ forms less than 1\% of the sulphur budget.  H$_2$S is still the dominant comet sulphur molecule, with SO$_2$ comprising $\sim$ 10\% of the sulphur budget. 

Qualitatively OCS is much lower in the comets than in IRAS 16293-2422B. Astrochemical models include the destruction of OCS by photons and cosmic rays producing CO + S or CS + O. OCS may also react with other atoms:
\begin{eqnarray}
\hbox{H} + \hbox{OCS} & \longrightarrow & \hbox{CO} + \hbox{HS}\\
\hbox{C} + \hbox{OCS} & \longrightarrow & \hbox{CO} + \hbox{CS}\\
\hbox{S} + \hbox{OCS} & \longrightarrow & \hbox{CO} + \hbox{S$_2$}
\end{eqnarray}
Subsequently S$_2$ could react with oxygen atoms forming SO, and HS with H to form H$_2$S. However, the exact pathways leading to the changes in composition seen in comets are still unclear.

\section{\label{sec:retention}To what degree do comets retain interstellar abundances?}

Figures~\ref{fig:common_coh_pie}-\ref{fig:common_s_pie} provide a  means to see how the different elements are partitioned among the various molecules in comets, how this partition varies from comet to comet, and between comets and the protostellar region. In general there is a wider range of molecules present in cometary ices compared to the protosellar gas (where the molecules are believed to result from evaporation of ices). More complex molecules are present and there is an evolution from the protostellar compositions for all three molecular families, with the dominant protostellar species being processed into other molecules, presumably during the disk/comet formation phase. None of the comets retain the same proportion of molecules as seen in the protostellar source. 
While some interstellar material may be included in each comet, disk-processed material must also be present. Here we consider whether there are any quantifiable methods to determine the relative degree to which comets contain material inherited from an earlier phase of evolution.  

\subsection{Coefficient of determination}
One way to determine the degree to which comets are related to the ISM
is to use the coefficient of determination, or $r^2$ value.   The $r^2$ value indicates the percentage of the comet composition that could be related to the molecules observed in IRAS 16293-2422B.  Although some care needs to be taken in this interpretation (it is valid only if the data and the fit are unbiased), we can use it as an indication of the degree to which  protostellar chemistry might determine the comet compositions. $r^2$ is defined as 
\begin{equation}
    r^2 = 1 - \frac{\sum (y_i - \hat{y}_i)^2}{\sum (y_i - \overline{y}_i)^2} \label{eq:r2}
\end{equation}
where $y_i$ is the comet data, $\hat{y}_i$ is the protostellar data and $\overline{y}_i$ is the average of the comet data.  $r^2$ is also the square of the Pearson correlation coefficient.

\begin{deluxetable}{lcccccc}
  \tablewidth{0pt}
    \tablecaption{\label{tab:inherit}Coefficient of determination ($r^2$) and regression  line slopes for the comparison of each molecular family with IRAS 16293-2422B for the set of common molecules listed in Table~\ref{tab:common}.  LS indicates slope calculated using the least-squares method, and ATS the slope from the Akritas-Thiel-Sen method.
    }
    \tablehead{
    \colhead{Comet} & \multicolumn{3}{c}{COH molecules} & \multicolumn{3}{c}{N molecules} \\
    & \colhead{$r^2$} & \multicolumn{2}{c}{line slope}     & \colhead{$r^2$} & \multicolumn{2}{c}{line slope}  \\
    \colhead{} & \colhead{} & \colhead{LS} & \colhead{ATS} & \colhead{} & \colhead{LS} & \colhead{ATS}
    }
\startdata
67P       & 0.956 & 0.716 & 0.719 & 0.741    & 0.798   & 1.219 \\ 
Hale-Bopp & 0.961 & 0.702 & 0.726 & 0.264    & 0.269   & 0.363\\
C/2014 Q2 & 0.914 & 0.690 & 0.672 & 0.641    & 0.466   & 0.495 \\
C/2021 A1 & 0.849 & 0.556 & 0.615 & 0.734    & 0.457   & 0.632 \\
C/2022 E3 & 0.841 & 0.550 & 0.531 & \nodata    & \nodata & \nodata\\
C/2013 R1 & 0.948 & 0.648 & 0.663 & \nodata & \nodata & \nodata
\enddata
\end{deluxetable}

We present $r^2$ values for the subset of comets with common molecule detections in Table~\ref{tab:inherit}.  Since the correlations are so weak for sulphur molecules we concentrate here on the COH and nitrogen-molecular families. For each comet $r^2$ is highest for the COH- molecules, with values ranging from 0.84 (for C/2022 E3) to 0.96 (for Hale-Bopp) suggesting that more than 84\% of the cometary composition can be explained by protostellar chemistry.  For N-molecules the $r^2$ values are lower and range from 0.26 (Hale-Bopp) to 0.74 (67P),  indicating between 26 and 73\% of the 
nitrogen molecules are inherited.

\subsection{Regression line slopes and deviation from protostellar compositions}

 An alternative way of looking at the change in comet composition compared to protostellar would be to look at the regression line slopes.  All of the cometary data can be fit by a straight line in log-log space ($x$(comet) = a $x$(PS)$^b$, where $x$(comet), $x$(PS) are the abundance ratios for the comet and protostellar region respectively and  $b$ is the slope of the fit). If there was perfect agreement (and therefore 100\% inheritance) between the comets and IRAS 16293-2422B then this would have a slope of 1. We argue that the deviation of the regression line slope from 1 could provide an indicator of the degree to which a comet contains disk processed material.

 We recalculate the regression line slopes for the comets sharing the common set of molecules listed in Table~\ref{tab:common}. We consider both the least-squares and ATS methods.  These updated slopes are listed in Table~\ref{tab:inherit}. For the COH- molecules we can order the comets from steepest to shallowest slope (least to most processed). Using ATS this is Hale-Bopp, 67P, C/2014 Q2,  C/2013 R1, C/2021 A1  and C/2022 E3. The same order is given by $r^2$ (see previous section) but is slightly different for the least-squares method which switches 67P and Hale-Bopp, and C/2014 Q2 and C/2013 R1.

In the case of nitrogen molecules we have included 67P.  However the ROSINA observations of this comet cannot distinguish between HCN and HNC, and there is only an upper limit for HNC in IRAS 16293-2422B.  This adds an additional uncertainty to the derived parameters for 67P.  Using ATS, 67P is again comet with least processing, followed by C/2014 Q2 and C/2021 A1, and finally Hale-Bopp being the most processed.
This is the same order as $r^2$, while the LTS regression line slopes switch C/2014 Q2 and C/2021 A1.

\section{\label{sec:disc}Discussion}

 The mixing ratios of the majority of the cometary molecules are within a factor of 10 of the protostellar values, indicating some degree of inheritance. Inheritance, though, cannot be the complete story. There are clear compositional differences between the comet population as a whole and IRAS 16293-2422B, in particular higher abundances of more complex molecules. If protostellar material is inherited by the disk, then disk processing must also have occurred to create these complex molecules. Experiments suggest that in the case of the COH and N families this is driven partly by UV/cosmic ray processing of the ices. While there is not equivalent experimental data for sulphur chemistry it seems reasonable to assume that this is also UV/cosmic ray driven.  The variation in mixing ratios observed among comets could be a result of formation from material processed under different irradiation environments. This might be achieved by formation at different locations and/or times during the disk evolution.  Mixing could also paly a role -- either vertical or radial mixing of icy grains would bring them into higher radiation environments and lead to chemical changes that might be preserved in comets.

The measurement of COM abundances in IRAS 16293-2422B may also be affected by differences in their binding energies, which result in varying emitting region sizes and can lead to an underestimation of their abundances relative to CH$_3$OH. \cite{frediani2025} recently presented the spatial distributions of COMs in IRAS 4A2, another hot corino. Their findings indicate that the detection of specific COMs depends on their distance from the protostar, with NH$_2$CHO and CH$_2$OHCHO located closest to the star, followed by CH$_3$CH$_2$OH, CH$_3$CHO, HCOOCH$_3$, and finally CH$_3$OH. Consequently, the emitting region for CH$_3$OH is larger than for the other COMs, causing their inferred abundances in the protostar to appear lower than they actually are. This may, in turn, lead to an overestimation of the inferred enhancement of their cometary abundances relative to protostellar values. However, there are also COMs (e.g., CH$_3$OCH$_3$, CH$_3$COCH$_3$, and C$_2$H$_5$CHO) whose binding energies are similar to or lower than that of CH$_3$OH and would therefore be detectable over the same or larger region of the protostellar source. These molecules have been detected in IRAS 16293-2422B and are found to be enhanced in our comets. While some care needs to be taken with COMs with higher binding energies, the change in the relative abundances of these molecules from protostellar region to comets lends further support to the idea that COMs are also formed within the disk.

The increase in cometary COMs was also found by \cite{bianchi19} who compared the abundance of COMs in comets 67P, Lemmon, Lovejoy, and Hale-Bopp with those detected in the hot corino SVS13-A. This is a source at a slightly later stage of evolution to IRAS 16293-2422B.  They found the abundance ratios of HCOOCH$_3$/CH$_3$OH and CH$_3$CH$_2$OH/CH$_3$OH were enhanced by less than a factor of 10, but the increase for NH$_2$CHO/CH$_3$OH and CH$_3$CHO/CH$_3$OH was $>$ 10.  Relative to IRAS 16293-2422B we see enhancements of $~\sim$ 10 for NH$_2$CHO/CH$_3$OH in the same comets - a slightly lower increase than seen relative to SVS13-A.  The abundance of COMs tends to be slightly higher in IRAS 16293-2422B than in SWS13-A (see Figure 4. of Bianchi et al. 2019). Hence the relative increase in COMs we see is less than they report. 

 \cite{lippi24} compared cometary H$_2$CO, CH$_3$OH and NH$_3$ with observations of protostellar disks. They used a statistical approach using the mean and standard deviation of the cometary population and found similar compositions for the different dynamical families, with good agreement for the ground-based infrared and sub(mm) observations.  They find that the ratio of CH$_3$OH/H$_2$CO is relative constant across interstellar clouds, protostellar regions and disks, with little variation found between comet dynamical families. These similarities suggest that the CH$_3$OH/H$_2$CO ratio is preserved during comet lifetimes, and lends support to the idea of inheritance, possibly by interstellar ices being enclosed in refractory grains until they are incorporated into comets. Inheritance is also supported by their analysis of the NH$_3$/H$_2$CO ratio which is consistent between comets and Class 0 and Class II disks.   This work is consistent with our analysis which also suggests a large degree of inheritance between protostellar regions and comets.

Statistically we find good correlations between cometary and protostellar abundance ratios for all three molecular families, although the correlations are better for the COH-family, for which a larger number of molecules are detected. 
A good question is how reliable are these correlations?  It might be expected that COMs are less abundant than simple molecules just because it takes longer for them to form.  Hence it might be expected that the Spearman's rank correlation coefficient, which measures the strength and direction of a monotonic relationship between two sets of variables might be high just because of this.  Additionally, uncertainties in the correlation coefficients can be very high when there is only a small sample of molecules, and caution is required in interpreting the results in these cases, e.g. in the case of the sulphur molecules.  We have attempted to address these concerns with the analysis in Sections~\ref{sec:trends} and \ref{sec:retention} by using the same set of molecules when considering trends in molecular abundance ratios in comets. But this is still unable to compare several potentially important carriers of carbon and oxygen, namely CO$_2$, C$_2$H$_2$, C$_2$H$_6$ and CH$_4$ which could not be detected by the ALMA survey of IRAS 16293-2422B.

For N-molecules, the abundances of NH$_3$ and HC$_3$N significantly impact the regression slope and statistical comparison in Figure~\ref{fig:n_ch3cn}. NH$_3$ is challenging to observe and consequently has not been detected in all comets. In our sample, upper limits are available for NH$_3$ in C/2021 A1 and 73P/SWS/B and C, while no detection is reported for C/2022 E3. As discussed in Section~\ref{sec:comet_obs} this could be due to observational contraints, rather than the absence of NH$_3$ in this comet.  Indeed,  given the widespread detection of NH$_3$ in other comets, it is likely present.
Additionally, the protostellar NH$_3$ abundance is  an upper limit.  If the actual abundance of NH$_3$ is much lower than assumed then this would affect the calculated correlation coefficients, although the slope of the regression line would still be strongly affect by the high HC$_3$N/CH$_3$CN in some comets.

 The HCN/NH$_3$ ratio appears consistently higher in comets. Again this effect maybe exaggerated because of the lack of an NH$_3$ detection in IRAS 16293-2422B. If the ratio really is higher in comets then it indicates conversion of NH$_3$ into HCN.

We have attempted to use statistical methods to estimate the relative degree of disk processing in comets. Using a subset of comets with a consistent set of detected molecules, we looked at how the $r^2$ value varies among comets. As an alternative method we also compared the regression line slopes, on the grounds that a larger variation from the slope of 1 line would indicate more processing and less inheritance. For the COH family both methods yield a consistent order for least to greatest disk processing of Hale-Bopp, 67P C/2013 R1, C/2014 Q2, C/2021 A1 and C/2022 E3, with only a minor  difference in the  least-squares regression results where C/2014 Q2 and C/2013 R1 are switched.

The relative ordering differs between the COH- and N-familes. While Hale-
Bopp appears to be least processed from its COH composition, it is most processed in its N molecules. Given that there is a larger number of detections of COH molecules the statistics and line fitting are likely to be more reliable for this family.

Given the similarities of the estimated relative processing for both $r^2$ and regression line slopes within a molecular family, we therefore suggest that both of these methods can be used to determine relative degree of inheritance, but only for a common set of molecules, and only for detections, without including upper limits.  However, we are not yet at a stage where either method can be used as a reliable way of differentiating between the degree of processing that interstellar material  has undergone to result in individual cometary compositions. To do this would require a much larger suite of observations for each comet covering molecules that are also observed in protostellar sources.  It would also be instructive to consider protostellar sources other than IRAS 16293-2422B.  There is no guarantee that IRAS 16293-2422B is compositionally similar to the protosolar nebula and it would be interesting to determine whether similar correlations are seen with other protostellar sources. 

There appear to be clear differences in the degree to which the different molecular families experience chemical processing in the disk.  One possibility is that the COH-molecules, such as CO, and CH$_3$COH - generally undergo less alteration between the protostellar phase and their incorporation into comets. This could be because these molecules form efficiently in the interstellar medium and freezeout onto icy dust grains at low temperatures without significant chemical modification. In contrast, nitrogen-bearing molecules such as NH$_3$, HCN and NH$_2$CHO undergo moderate processing due to reactions that require higher temperatures or energetic inputs. For example, HCN forms in the gas phase in warm regions, but can also be altered by photodissociation.  

\section{\label{sec:conclusions}Conclusions}

Our work suggests that all comets contain disk-processed material to some extent, as well as some amount of inherited interstellar material. Evaluating the degree to which this is the case is difficult because of the lack of a number of comets in which the same set of molecules has been observed. Our analysis therefore is preliminary and awaits the development of a larger, more consistent comet database. However, it provides a potential framework for analyzing the relative contribution of interstellar and disk molecular to comet compositions with the potential of characterizing the formation conditions of individual comets or cometary families.

The full dataset considered in our study includes a significant number of upper limits, which can introduce challenges in statistical analysis. Ignoring them means leaving out potentially important data, including molecules which incorporate high fractions of particular elements, e.g. NH$_3$. Including them as a fixed value (e.g.\ 50\% of the reported upper limit) may also be unreliable.  We attempt to mitigate the problems caused by upper limits by using the Kendall correlation coefficients and the ATS method to determine regression line slopes. 

To determine the relative degree of processing among comets We also considered a smaller subset of comets in which the same molecules have {\it measured} abundances.  While ignoring some important molecules this does have the advantage of allowing a qualitative comparison between comets.

Our conclusions can be summarized as follows:

\begin{enumerate}
\item The composition of all of the comets in our sample show a good correlation with the protostellar composition, as determined using Pearson correlation coefficients and Spearman's rank correlation.

    \item The different molecular families show different degrees of correlation. COH- molecule abundance ratios show the tightest degree of correlation with protostellar observations, and this does not depend on the reference molecule used. This suggests that  chemical processing in the disk  increases the abundance of more complex COH-molecules such as HCOOCH$_3$, CH$_3$CHO, and NH$_2$CHO. The majority of this family of molecules is inherited from the protostellar region ($\gtrapprox$ 90\%).
    \item N-molecules are less tightly correlated with protostellar composition than COH-molecules.  The coefficient of determination suggests around 60 -- 77\% of the nitrogen molecular mixing ratios relative to CH$_3$CN may be accounted for by inheritance from the protostellar region. Disk processing is needed to account for the increase in HCN and HC$_3$N seen in comets, and the decrease in NH$_3$.
    \item  For most comets there is only weak correlation between their sulphur chemistry and that of the protostar. The comparison for this family is hampered by the relatively low number of detections and the high number of upper limits.  More observational data may in the future reveal correlations for this molecular family, similar to those seen for the COH- and N- molecules.
    \item  With a common set of molecular {\it detections} from the COH-molecular family we find that comets can tentatively be ordered according to the degree to which their composition is inherited from the interstellar medium.  There is reasonable consistency between the order determine from the $r^2$ value and from the regression line slope for the COH and N molecular families.  Based on this it may be possible, if a large number of common observations can be made for each comet, for this to be extended to other comets and to other molecular families, raising the possibility that we may in the future be able to relate comet compositions to their time and location of formation in the solar nebula.
\end{enumerate}

\acknowledgments
This research was conducted at the Jet Propulsion Laboratory,
California Institute of Technology under contract with the National 
Aeronautics and Space Administration (80NM0018D004).  Support for K.W. was provided by NASA/Emerging Worlds Program (22-EW22-0033).
 L.M. acknowledges financial support from DAE and DST-SERB research grant (MTR/2021/000864) from the Government of India for this work. E.G. acknowledges financial support from the NSF Astronomy and Astrophysics program (2009910).

 © 2024. All rights reserved.

\software{The ATS regression lines and the correlation coefficients were calculated in R studio (version 2025.05.1+513) \citep{rstudio} using the \texttt{cenken} function from the NADA2 ''R" package \citep{NADA2}. The least-squares regression lines were calculated using IDL\textsuperscript{\textregistered} (version 8.4). }

\bibliography{disk}

\end{document}